\documentstyle[preprint,aps,epsfig,floats]{revtex}
\tightenlines
\newcommand{\figsize}{140mm}

\begin{document}

\renewcommand{\d}{\mbox{d}}
\newcommand{\ie}{\mbox{\rm i.e.}}
\newcommand{\eg}{\mbox{\rm e.g.}}
\newcommand{\ifm}[1]{\relax\ifmmode #1\else $#1$\fi}
\newcommand{\etal}{{\it et al.}}
\newcommand{\Dzero}{D\O}
\newcommand{\delphi}{DELPHI Collaboration}
\newcommand{\uaone}{UA1 Collaboration}
\newcommand{\uatwo}{UA2 Collaboration}
\newcommand{\cdf}{CDF Collaboration}
\newcommand{\dzero}{\Dzero\ Collaboration}
\newcommand{\alephc}{ALEPH Collaboration}
\newcommand{\Lthree}{L3 Collaboration}
\newcommand{\opalc}{OPAL Collaboration}
\newcommand{\PRL}{Phys. Rev. Lett.}
\newcommand{\PL}{Phys. Lett.}
\newcommand{\PR}{Phys. Rev.}
\newcommand{\NP}{Nucl. Phys.}
\newcommand{\NIM}{Nucl. Instrum. Methods in Phys. Res.}
\newcommand{\ZP}{Z.~Phys.}
\newcommand{\GEAN}{{\footnotesize GEANT}}
\newcommand{\ISAJ}{{\footnotesize ISAJET}}
\newcommand{\HERW}{{\footnotesize HERWIG}}
\newcommand{\PDFL}{{\footnotesize PDFLIB}}
\newcommand{\PYTH}{{\footnotesize PYTHIA}}
\newcommand{\JETR}{{\footnotesize JETRAD}}
\newcommand{\RECO}{{\footnotesize \Dzero RECO}}

\newcommand{\VTX}{VTX}
\newcommand{\CDC}{CDC}
\newcommand{\FDC}{FDC}
\newcommand{\TRD}{TRD}
\newcommand{\CC}{CC}
\newcommand{\EC}{EC}
\newcommand{\EM}{EM}
\newcommand{\HAD}{HAD}
\newcommand{\dedx}{\ifm{dE/dx}}
\newcommand{\lzero}{L\O}
\newcommand{\alphas}{\ifm{\alpha_s}}
\newcommand{\sqrts}{\ifm{\sqrt{s}}}
\newcommand{\abseta}{\ifm{|\eta|}}
\newcommand{\absetad}{\ifm{|\eta_D|}}
\newcommand{\etad}{\ifm{\eta_D}}
\newcommand{\absz}{\ifm{|z|}}
\newcommand{\absdphi}{\ifm{|\Delta\phi|}}
\newcommand{\qt}{\ifm{q_T}}
\newcommand{\px}{\ifm{p_x}}
\newcommand{\py}{\ifm{p_y}}
\newcommand{\pz}{\ifm{p_z}}
\newcommand{\mt}{\ifm{m_T}}
\newcommand{\et}{\ifm{E_T}}
\newcommand{\etv}{\ifm{\vec\et}}
\newcommand{\met}{\mbox{${\hbox{$E$\kern-0.6em\lower-.1ex\hbox{/}}}_T$}}
\newcommand{\pt}{\ifm{p_T}}
\renewcommand{\pt}{\ifm{p_T}}
\newcommand{\ptv}{\ifm{\vec\pt}}
\newcommand{\mpt}{\mbox{$\rlap{\kern0.1em/}\pt$}}
\newcommand{\mptv}{\mbox{$\rlap{\kern0.1em/}\vec\pt$}}
\newcommand{\PM}{\ifm{\pm}}
\newcommand{\lt}{\ifm{<}}
\newcommand{\gt}{\ifm{>}}
\newcommand{\ptg}{\ifm{\pt(\gamma)}}
\newcommand{\Eg}{\ifm{E(\gamma)}}
\newcommand{\phig}{\ifm{\phi(\gamma)}}
\newcommand{\etag}{\ifm{\eta(\gamma)}}
\newcommand{\uv}{\ifm{\vec u}}
\newcommand{\ut}{\ifm{u_T}}
\newcommand{\utv}{\ifm{\vec\ut}}
\newcommand{\upar}{\ifm{u_{\parallel}}}
\newcommand{\uper}{\ifm{u_{\perp}}}
\newcommand{\phier}{\ifm{\phi(e,R)}}
\newcommand{\ueta}{\ifm{u_\eta}}
\newcommand{\uxi}{\ifm{u_\xi}}
\newcommand{\phir}{\ifm{\phi(R)}}
\newcommand{\sigmaw}{\ifm{\sigma(p \overline{p} \rightarrow W+X) 
\cdot B(W \rightarrow e \nu)}}
\newcommand{\bwev} {\ifm{B(W \rightarrow e \nu)}}
\newcommand{\wb}{\ifm{W}}
\newcommand{\wbp}{\ifm{W^+}}
\newcommand{\wbm}{\ifm{W^-}}
\newcommand{\wev}{\ifm{W\to e\nu}}
\newcommand{\wegam}{\ifm{\wev\gamma}}
\newcommand{\wtv}{\ifm{W\to \tau\nu}}
\newcommand{\wte}{\ifm{W\to \tau\nu\to e\nu\overline\nu\nu}}
\newcommand{\wth}{\ifm{W\to \tau\nu\to \hbox{hadrons}+X}}
\newcommand{\dsigmadptdyw}{\ifm{d^2\sigma_{W}/d\pt dy}}
\newcommand{\mw}{\ifm{M_{W}}}
\newcommand{\wwidth}{\ifm{\Gamma_{W}}}
\newcommand{\ptw}{\ifm{\pt(W)}}
\newcommand{\ptwv}{\ifm{\vec\ptw}}
\newcommand{\sigmaz}{\ifm{\sigma(p \overline{p} \rightarrow Z+X)
 \cdot B(Z \rightarrow ee)}}
\newcommand{\zb}{\ifm{Z}}
\newcommand{\zee}{\ifm{Z\to ee}}
\newcommand{\ztt}{\ifm{Z\to \tau \tau}}
\newcommand{\zeegam}{\ifm{\zee\gamma}}
\newcommand{\dsigmadptdyz}{\ifm{d^2\sigma_{Z}/d\pt dy}}
\newcommand{\mz}{\ifm{M_{Z}}}
\newcommand{\ptz}{\ifm{\pt(Z)}}
\newcommand{\dphiee}{\ifm{\Delta\phi(ee)}}
\newcommand{\ptee}{\ifm{\pt(ee)}}
\newcommand{\pteev}{\ifm{\vec\ptee}}
\newcommand{\pee}{\ifm{p(ee)}}
\newcommand{\peev}{\ifm{\vec\pee}}
\newcommand{\mee}{\ifm{m(ee)}}
\newcommand{\pteta}{\ifm{p_\eta (ee)}}
\newcommand{\etabalcor}{\ifm{\left(\utv/\rrec+\pteev\right)\cdot\hat{\eta}}}
\newcommand{\etabal}{\ifm{\left(\utv+\pteev\right)\cdot\hat{\eta}}}
\newcommand{\xibalcor}{\ifm{\left(\utv/\rrec+\pteev\right)\cdot\hat{\xi}}}
\newcommand{\xibal}{\ifm{\left(\utv+\pteev\right)\cdot\hat{\xi}}}
\newcommand{\Reqn}{\ifm{\sigma(p \overline{p} \rightarrow W+X) 
\cdot B(W \rightarrow e \nu)
/ \sigma(p \overline{p} \rightarrow Z+X) \cdot B(Z \rightarrow ee)}}
\newcommand{\R}{\ifm{R}}
\newcommand{\tev}{\ifm{\tau\to e\nu\overline\nu}}
\newcommand{\qqbar}{\ifm{q\overline{q}}}
\newcommand{\bbbar}{\ifm{b\overline{b}}}
\newcommand{\ee}{\ifm{e^+e^-}}
\newcommand{\ppbar}{\ifm{p\overline{p}}}
\newcommand{\ppbarw}{\ifm{\ppbar\to\wb}}
\newcommand{\ppbarwp}{\ifm{\ppbar\to\wbp}}
\newcommand{\ppbarwev}{\ifm{\ppbar\to\wev}}
\newcommand{\ppbarwevp}{\ifm{\ppbar\to W^+ \to e^+\nu}}
\newcommand{\TeV}{\mbox{TeV}}
\newcommand{\GeV}{\mbox{GeV}}
\newcommand{\GeVm}{\mbox{GeV}}
\newcommand{\GEVM}{\mbox{GeV/$c^2$}}
\newcommand{\GEVP}{\mbox{GeV/$c$}}
\newcommand{\MeV}{\mbox{MeV}}
\newcommand{\MeVp}{\mbox{MeV}}
\newcommand{\MeVm}{\mbox{MeV}}
\newcommand{\MEVP}{\mbox{MeV/$c$}}
\newcommand{\MEVM}{\mbox{MeV/$c^2$}}
\newcommand{\inb}{\mbox{nb$^{-1}$}}
\newcommand{\ipb}{\mbox{pb$^{-1}$}}
\newcommand{\nb}{\mbox{nb}}
\newcommand{\pb}{\mbox{pb}}
\newcommand{\Ee}{\ifm{E(e)}}
\newcommand{\pex}{\ifm{\px(e)}}
\newcommand{\pey}{\ifm{\py(e)}}
\newcommand{\pez}{\ifm{\pz(e)}}
\newcommand{\pe}{\ifm{p(e)}}
\newcommand{\pev}{\ifm{\vec\pe}}
\newcommand{\pte}{\ifm{\pt(e)}}
\newcommand{\ptev}{\ifm{\vec\pte}}
\newcommand{\peonev}{\ifm{\vec p(e_1)}}
\newcommand{\petwov}{\ifm{\vec p(e_2)}}
\newcommand{\phie}{\ifm{\phi(e)}}
\newcommand{\te}{\ifm{\theta(e)}}
\newcommand{\etae}{\ifm{\eta(e)}}
\newcommand{\absetae}{\ifm{|\etae|}}
\newcommand{\rcal}{\ifm{r_{\rm cal}}}
\newcommand{\rvtx}{\ifm{r_{\rm vtx}}}
\newcommand{\rtrk}{\ifm{r_{\rm trk}}}
\newcommand{\xcal}{\ifm{x_{\rm cal}}}
\newcommand{\ycal}{\ifm{y_{\rm cal}}}
\newcommand{\zcal}{\ifm{z_{\rm cal}}}
\newcommand{\xtrk}{\ifm{x_{\rm trk}}}
\newcommand{\ytrk}{\ifm{y_{\rm trk}}}
\newcommand{\ztrk}{\ifm{z_{\rm trk}}}
\newcommand{\xvtx}{\ifm{x_{\rm vtx}}}
\newcommand{\yvtx}{\ifm{y_{\rm vtx}}}
\newcommand{\zvtx}{\ifm{z_{\rm vtx}}}
\newcommand{\fiso}{\ifm{f_{\rm iso}}}
\newcommand{\emfr}{\ifm{f_{\rm EM}}}
\newcommand{\sigm}{\ifm{\sigma_{\rm trk}}}
\newcommand{\hmtx}{\ifm{\chi^2_{hmtx}}}
\newcommand{\Enu}{\ifm{E(\nu)}}
\newcommand{\pnux}{\ifm{\px(\nu)}}
\newcommand{\pnuy}{\ifm{\py(\nu)}}
\newcommand{\pnuz}{\ifm{\pz(\nu)}}
\newcommand{\pnu}{\ifm{p(\nu)}}
\newcommand{\pnuv}{\ifm{\vec\pnu}}
\newcommand{\ptnu}{\ifm{\pt(\nu)}}
\newcommand{\ptnuv}{\ifm{\vec\ptnu}}
\newcommand{\phinu}{\ifm{\phi(\nu)}}
\newcommand{\rem}{\ifm{{\rm R_{EM}}}}
\newcommand{\rrec}{\ifm{{\rm R_{rec}}}}
\newcommand{\sigrec}{\ifm{\sigma_{\rm rec}}}
\newcommand{\dupar}{\ifm{\Delta \upar}}
\newcommand{\mdupar}{\ifm{\langle \dupar \rangle}}
\newcommand{\alphaem}{\ifm{\alpha_{\rm EM}}}
\newcommand{\deltaem}{\ifm{\delta_{\rm EM}}}
\newcommand{\alphamb}{\ifm{\alpha_{\rm mb}}}
\newcommand{\srec}{\ifm{s_{\rm rec}}}
\newcommand{\alpharec}{\ifm{\alpha_{\rm rec}}}
\newcommand{\betarec}{\ifm{\beta_{\rm rec}}}
\newcommand{\gammin}{\ifm{\gamma_{\rm min}}}
\newcommand{\regam}{\ifm{\Delta R(e\gamma)}}
\newcommand{\rzero}{\ifm{R_0}}
\newcommand{\rcoal}{\ifm{{\rm R}_{\rm coalesce}}}
\newcommand{\fss}{\ifm{f_{\rm ss}}}
\newcommand{\alphacdc}{\ifm{\alpha_{\rm CDC}}}
\newcommand{\betacdc}{\ifm{\beta_{\rm CDC}}}
\newcommand{\alphacc}{\ifm{\alpha_{\rm CC}}}
\newcommand{\gtwo}{\ifm{g_2}}
\newcommand{\rl}{\ifm{X_0}}
\newcommand{\il}{\ifm{\lambda_A}}
\newcommand{\cem}{\ifm{c_{\rm EM}}}
\newcommand{\sem}{\ifm{s_{\rm EM}}}
\newcommand{\nem}{\ifm{n_{\rm EM}}}
\newcommand{\lqcd}{\ifm{\Lambda_{\rm QCD}}}


\newcommand{\wxsec}{2310}
\newcommand{\wxstat}{10}
\newcommand{\wxsyst}{50}
\newcommand{\wxlum}{100}
\newcommand{\werr}{110}        
\newcommand{\wnum}{67078}
\newcommand{\wacc}{0.465}
\newcommand{\waccerr}{0.004}
\newcommand{\weff}{0.671}
\newcommand{\wefferr}{0.009}
\newcommand{\wfqcd}{0.064}
\newcommand{\wfqcderr}{0.014}
\newcommand{\wfqcdcc}{0.046}
\newcommand{\wfqcdccerr}{0.014}
\newcommand{\wfqcdec}{0.143}
\newcommand{\wfqcdecerr}{0.043}
\newcommand{\wacczinw}{0.133}
\newcommand{\wacczinwerr}{0.034}
\newcommand{\wacctau}{0.021}
\newcommand{\wacctauerr}{0.002}
\newcommand{\wtaucorr}{1.021}


\newcommand{\zxsec}{221}
\newcommand{\zxstat}{3}
\newcommand{\zxsyst}{4}
\newcommand{\zxlum}{10}
\newcommand{\zerr}{11}         
\newcommand{\znum}{5397}
\newcommand{\zacc}{0.366}
\newcommand{\zaccerr}{0.003}
\newcommand{\zeff}{0.744}
\newcommand{\zefferr}{0.011}
\newcommand{\zfdy}{0.012}
\newcommand{\zfdyerr}{0.001}
\newcommand{\zfqcd}{0.045}
\newcommand{\zfqcderr}{0.005}


\newcommand{\lumb}{84.5}
\newcommand{\lumberr}{3.6}
\newcommand{\luma}{13}
\newcommand{\lumab}{97.5}
\newcommand{\lumamu}{11}


\newcommand{\rxsec}{10.43}
\newcommand{\rxstat}{0.15}
\newcommand{\rxsyst}{0.20}
\newcommand{\rxnlo}{0.10}
\newcommand{\rerr}{0.27}       
\newcommand{\rnum}{12.43}
\newcommand{\rnumerr}{0.18}
\newcommand{\racc}{0.787}
\newcommand{\raccerr}{0.007}
\newcommand{\reff}{1.108}
\newcommand{\refferr}{0.007}
\newcommand{\rerrwqcd}{1.5\%}
\newcommand{\rerrzstat}{1.4\%}
\newcommand{\rerracc}{0.8\%}
\newcommand{\rerreff}{0.6\%}
\newcommand{\rerrzqcd}{0.5\%}
\newcommand{\rerrnlo}{1.0\%}


\newcommand{\brwev}{0.1066}
\newcommand{\brstat}{0.0015}
\newcommand{\brsyst}{0.0021}
\newcommand{\brthy}{0.0011}
\newcommand{\brnlo}{0.0011}
\newcommand{\brerr}{0.0030}     


\newcommand{\gw}{2.130}
\newcommand{\gwstat}{0.030}
\newcommand{\gwsyst}{0.041}
\newcommand{\gwthy}{0.022}
\newcommand{\gwnlo}{0.021}
\newcommand{\gwerr}{0.060}      


\newcommand{\gwinvmev}{168} 
\newcommand{\gwinvgev}{0.168}


\newcommand{\wxssix}{658}
\newcommand{\wxssixstat}{58}
\newcommand{\wxssixsys}{34}    
\newcommand{\wxssixerr}{67}    
\newcommand{\effsix}{0.799}
\newcommand{\effsixerr}{0.024}
\newcommand{\accsix}{0.521}
\newcommand{\accsixerr}{0.013}
\newcommand{\fqcdsix}{0.016}
\newcommand{\fqcdsixerr}{0.012}
\newcommand{\sslum}{505}
\newcommand{\sslumerr}{15}
\newcommand{\ssnw}{130}
\newcommand{\ssnwcc}{119}
\newcommand{\ssnwec}{11}
\newcommand{\sslzeff}{0.897}
\newcommand{\sslzefferr}{0.009}
\newcommand{\lzeight}{0.905}
\newcommand{\lzsix}{0.823}
\newcommand{\sspj}{1.5}
\newcommand{\ssrs}{0.28}
\newcommand{\ssrserr}{0.11}
\newcommand{\ssnzeight}{621}
\newcommand{\ssnzsix}{1.2}
\newcommand{\ssnzsixerr}{0.3}


\newcommand{\nlow}{0.998}
\newcommand{\nlowerr}{0.001}
\newcommand{\nloz}{1.005}
\newcommand{\nlor}{1.00}
\newcommand{\nlorerr}{0.01}    


\newcommand{\ra}{10.82}
\newcommand{\raerrcorr}{0.141}
\newcommand{\raerruncorr}{0.408}

\newcommand{\ramu}{11.8}
\newcommand{\ramuerruncorr}{2.110}

\newcommand{\rberrcorr}{0.141}
\newcommand{\rberruncorr}{0.235}

\newcommand{\rcomb}{10.54}
\newcommand{\rcomberr}{0.24}

\newcommand{\gwcomb}{2.107}
\newcommand{\gwcomberr}{0.054}

\newcommand{\brcomb}{0.108}
\newcommand{\brcomberr}{0.003}

\newcommand{\gwinvcomb}{0.132}


\title{Extraction of the Width of the $W$ Boson from
 Measurements of \sigmaw\ and \sigmaz\ 
and their Ratio}

%
\author{                                                                      
B.~Abbott,$^{45}$                                                             
M.~Abolins,$^{42}$                                                            
V.~Abramov,$^{18}$                                                            
B.S.~Acharya,$^{11}$                                                          
I.~Adam,$^{44}$                                                               
D.L.~Adams,$^{54}$                                                            
M.~Adams,$^{28}$                                                              
S.~Ahn,$^{27}$                                                                
V.~Akimov,$^{16}$                                                             
G.A.~Alves,$^{2}$
% ...
\\                                                                            
\vskip 0.30cm                                                                 
\centerline{(D\O\ Collaboration)}                                             
\vskip 0.30cm                                                                 
}                                                                             
\address{                                                                     
\centerline{$^{1}$Universidad de Buenos Aires, Buenos Aires, Argentina}       
\centerline{$^{2}$LAFEX, Centro Brasileiro de Pesquisas F{\'\i}sicas,         
                  Rio de Janeiro, Brazil}                                     
\centerline{$^{3}$Universidade do Estado do Rio de Janeiro,                   
                  Rio de Janeiro, Brazil}                                     
\centerline{$^{4}$Institute of High Energy Physics, Beijing,                  
                  People's Republic of China}                                 
\centerline{$^{5}$Universidad de los Andes, Bogot\'{a}, Colombia}             
\centerline{$^{6}$Universidad San Francisco de Quito, Quito, Ecuador}         
\centerline{$^{7}$Institut des Sciences Nucl\'eaires, IN2P3-CNRS,             
                  Universite de Grenoble 1, Grenoble, France}                 
\centerline{$^{8}$DAPNIA/Service de Physique des Particules, CEA, Saclay,     
                  France}                                                     
\centerline{$^{9}$Panjab University, Chandigarh, India}                       
\centerline{$^{10}$Delhi University, Delhi, India}                            
\centerline{$^{11}$Tata Institute of Fundamental Research, Mumbai, India}     
\centerline{$^{12}$Kyungsung University, Pusan, Korea}                        
\centerline{$^{13}$Seoul National University, Seoul, Korea}                   
\centerline{$^{14}$CINVESTAV, Mexico City, Mexico}                            
\centerline{$^{15}$Institute of Nuclear Physics, Krak\'ow, Poland}            
\centerline{$^{16}$Institute for Theoretical and Experimental Physics,        
                   Moscow, Russia}                                            
\centerline{$^{17}$Moscow State University, Moscow, Russia}                   
\centerline{$^{18}$Institute for High Energy Physics, Protvino, Russia}       
\centerline{$^{19}$Lancaster University, Lancaster, United Kingdom}           
\centerline{$^{20}$University of Arizona, Tucson, Arizona 85721}              
\centerline{$^{21}$Lawrence Berkeley National Laboratory and University of    
                   California, Berkeley, California 94720}                    
\centerline{$^{22}$University of California, Davis, California 95616}         
\centerline{$^{23}$University of California, Irvine, California 92697}        
\centerline{$^{24}$University of California, Riverside, California 92521}     
\centerline{$^{25}$Florida State University, Tallahassee, Florida 32306}      
\centerline{$^{26}$University of Hawaii, Honolulu, Hawaii 96822}              
\centerline{$^{27}$Fermi National Accelerator Laboratory, Batavia,            
                   Illinois 60510}                                            
\centerline{$^{28}$University of Illinois at Chicago, Chicago,                
                   Illinois 60607}                                            
\centerline{$^{29}$Northern Illinois University, DeKalb, Illinois 60115}      
\centerline{$^{30}$Northwestern University, Evanston, Illinois 60208}         
\centerline{$^{31}$Indiana University, Bloomington, Indiana 47405}            
\centerline{$^{32}$University of Notre Dame, Notre Dame, Indiana 46556}       
\centerline{$^{33}$Purdue University, West Lafayette, Indiana 47907}          
\centerline{$^{34}$Iowa State University, Ames, Iowa 50011}                   
\centerline{$^{35}$University of Kansas, Lawrence, Kansas 66045}              
\centerline{$^{36}$Kansas State University, Manhattan, Kansas 66506}          
\centerline{$^{37}$Louisiana Tech University, Ruston, Louisiana 71272}        
\centerline{$^{38}$University of Maryland, College Park, Maryland 20742}      
\centerline{$^{39}$Boston University, Boston, Massachusetts 02215}            
\centerline{$^{40}$Northeastern University, Boston, Massachusetts 02115}      
\centerline{$^{41}$University of Michigan, Ann Arbor, Michigan 48109}         
\centerline{$^{42}$Michigan State University, East Lansing, Michigan 48824}   
\centerline{$^{43}$University of Nebraska, Lincoln, Nebraska 68588}           
\centerline{$^{44}$Columbia University, New York, New York 10027}             
\centerline{$^{45}$New York University, New York, New York 10003}             
\centerline{$^{46}$University of Rochester, Rochester, New York 14627}        
\centerline{$^{47}$State University of New York, Stony Brook,                 
                   New York 11794}                                            
\centerline{$^{48}$Brookhaven National Laboratory, Upton, New York 11973}     
\centerline{$^{49}$Langston University, Langston, Oklahoma 73050}             
\centerline{$^{50}$University of Oklahoma, Norman, Oklahoma 73019}            
\centerline{$^{51}$Brown University, Providence, Rhode Island 02912}          
\centerline{$^{52}$University of Texas, Arlington, Texas 76019}               
\centerline{$^{53}$Texas A\&M University, College Station, Texas 77843}       
\centerline{$^{54}$Rice University, Houston, Texas 77005}                     
}                                                                             

\maketitle

\vspace{0.5in}
\begin{abstract}
We report on measurements of inclusive cross sections times
branching fractions into electrons 
for $W$ and $Z$ bosons produced in \ppbar\
collisions at \sqrts\ = 1.8~\TeV. From an integrated luminosity of
\lumb~\ipb\ recorded in 1994--1995 using the \Dzero\ detector at the
Fermilab Tevatron,  we determine
\sigmaw = \wxsec\ $\pm$ \wxstat\ (stat) $\pm$ \wxsyst\ (syst) $\pm$ 
\wxlum\ (lum) \pb\ and
\sigmaz = \zxsec\ $\pm$ \zxstat\ (stat) $\pm$ \zxsyst\ (syst) $\pm$ 
\zxlum\ (lum) \pb.
From these, we derive \Reqn\ = \rxsec\ $\pm$ \rxstat\ (stat) $\pm$ 
\rxsyst\ (syst) $\pm$ \rxnlo\ (NLO), \bwev\ = \brwev\ $\pm$ \brstat\ (stat) 
$\pm$ \brsyst\ (syst) $\pm$ \brthy\ (theory) $\pm$ \brnlo\ (NLO), and 
$\Gamma_W$ = \gw\ $\pm$ \gwstat\ (stat) $\pm$
\gwsyst\ (syst) $\pm$ \gwthy\ (theory) $\pm$ \gwnlo\ (NLO) \GeV.  
We use the latter to set a 95\% confidence level
upper limit on the partial decay width of the \wb\ boson into
non-standard model final states, $\Gamma_W^{inv}$, of \gwinvgev\ \GeV.  
Combining these results with those from the 1992--1993 data gives 
\Reqn\ = \rcomb\ $\pm$ \rcomberr, 
$\Gamma_W$ = \gwcomb\ $\pm$ \gwcomberr\ \GeV, and a 95\% C.L. upper limit on 
$\Gamma_W^{inv}$ of \gwinvcomb\ \GeV.
Using a sample with a
luminosity of 505 \inb\ taken at \sqrts\ = 630 \GeV, we measure
\sigmaw\ = \wxssix\ $\pm$ \wxssixerr\ \pb .
 

\end{abstract}




\section{Introduction}
\label{sec:intro}
Since their discovery in
1983~\cite{wdiscovery}, comparison of the properties of $W$ and $Z$ bosons
to predictions of the standard model has been a subject of intense
study~\cite{Wpapers,ua1r,ua2r,cdf1aR,d01aR,LEPwwidth}. One
such property is the $W$ boson width.
Within the standard model, the $W$ boson decays into 
quark or lepton electroweak
doublets.  To lowest order, the partial decay width of the $W$ boson 
into massless fermions $f \bar{f'}$  can be written as
\begin{equation}
\Gamma_{W\rightarrow f \bar{f'}} = |V_{f \bar{f'}}|^2
N_C(G_F/\sqrt{2}) (M_W^3/6 \pi)
\end{equation}
where $V_{f \bar{f'}}$ are the Kobayashi-Maskawa matrix elements for
decays into quarks and unity for decays into leptons. The term $N_C$ 
accounts for color and is $3 (1 +
\alpha_s(M_W)/\pi + \ldots$) for decays into quarks and unity for leptonic
decays.  Within the standard model, the total width of the $W$ boson is 
the sum of the partial widths
over three generations of lepton doublets and two generations of
quark doublets.
If additional non-standard model particles exist, which are
lighter than and couple to the \wb\ boson, then the width would have an
additional contribution. One example is a
supersymmetric model in which the \wb\ boson can decay to the lightest
super-partner of the charged gauge bosons and the lightest
super-partner of the neutral gauge bosons, with a width that
depends on the masses of the super-particles~\cite{susywidth}. Thus, the \wb\  
boson width is of interest as a test of the standard model 
and as a probe for new physics.

The \wb\ boson width has been measured indirectly by the
UA1~\cite{ua1r}, UA2~\cite{ua2r}, CDF~\cite{cdf1aR},
and \Dzero~\cite{d01aR} collaborations.  The most recent
results are $\Gamma_W = 2.044 \pm 0.093$ \GeV\ from \Dzero\ and
$\Gamma_W = 2.064 \pm 0.084$ \GeV\ from CDF.
Both used a method which is based on
measuring  the ratio ${\cal R}$ of the \wev\ and
\zee\ cross sections:    
\begin{equation}
{\cal R} \equiv
\frac
{\sigma(p \overline{p} \rightarrow W+X) \cdot B(W \rightarrow e \nu)}
{\sigma(p \overline{p} \rightarrow Z+X) \cdot B(Z \rightarrow ee)}.
\end{equation}
The width can be calculated from this measurement using
\begin{equation}
{\cal R}  =
          \frac{\sigma_W}{\sigma_Z}\cdot
               \frac{\Gamma_Z}{\Gamma_{Z\to ll}}\cdot
               \frac{\Gamma_{W\to l \nu}}{\Gamma_W}.
\end{equation}
Both
$\sigma_W / \sigma_Z$ and $\Gamma_{W\to l \nu}$ can be
calculated theoretically to high precision~\cite{theoryrs},  
and depend only on the couplings of the $W$ and
$Z$ bosons to the lepton and quark doublets, and the ratio $\Gamma_Z /
\Gamma_{Z\to ll}$ has been measured  precisely by
experiments at LEP~\cite{lep_zmeasurements}.

The $W$ boson width has also been measured by the L3 and OPAL collaborations
at LEP~\cite{LEPwwidth} using kinematic fits to $qqqq$ and $qql\nu$ events,  
and by CDF~\cite{CDFwwidth} by looking at the high-mass tail of the transverse 
mass spectrum. Their current results are $\Gamma_W = 1.97 \pm 0.38 $ \GeV,
$\Gamma_W = 1.84 \pm 0.38 $ \GeV, and $\Gamma_W = 2.11 \pm 0.32\ $ \GeV, 
respectively. 
 
This paper presents new measurements of \sigmaw, \sigmaz, and their ratio 
$\cal R$ using a data sample approximately six times larger than was used in 
the previous \Dzero\ measurements. The value of $\cal{R}$ is used to extract 
the branching fraction $B(W \rightarrow e \nu)$ and the total decay width of 
the \wb\ boson, $\Gamma_W$.  
We set an upper limit on the partial decay width 
of the \wb\ boson to states not included in the standard model.

The uncertainties on the measurements of the absolute cross sections
are dominated by
the uncertainty on the integrated luminosity measurement (4.3\%). 
In the ratio, many of the systematic uncertainties, including that on
luminosity, cancel.  The uncertainty in $\cal{R}$ is dominated by 
the uncertainty in the QCD background 
in the \wb\ boson sample (\rerrwqcd); 
the statistics of the $Z$ boson sample
(\rerrzstat);
the uncertainty in the ratio of the acceptances for $W$ and $Z$ bosons
(\rerracc); 
the uncertainty in the ratio of the electron identification efficiencies 
for $W$ and $Z$ bosons (\rerreff); 
and the uncertainty in the multijet, $b$ quark, and direct photon backgrounds 
to the \zb\ boson (\rerrzqcd). In addition, we assign a 1\% theoretical 
uncertainty 
on $\cal R$ due to next-to-leading-order electroweak radiative corrections.

The paper is organized as follows.
Section~\ref{sec:detector} is a brief description of the \Dzero\ 
detector, emphasizing the components important for this analysis.
Section~\ref{sec:dataselection} describes the criteria used to select the 
\wev\ and \zee\ data samples.  
Section~\ref{sec:acceptance} describes the calculation
of the kinematic and geometric acceptance for the selection criteria. 
Section~\ref{sec:efficiencies} presents the measurement of the electron 
identification efficiency.  
Section~\ref{sec:backgrounds} presents the estimate of
the  backgrounds in the data samples.
Section~\ref{sec:luminosity} gives some details about the luminosity 
measurement.
Sections~\ref{sec:crosssectionresults} and~\ref{sec:checks} present the 
cross section results and some consistency checks, respectively.
Section~\ref{sec:630} presents the measurement of the $W$ boson cross section
times branching fraction into electrons at \sqrts\ = 630~\GeV.
Section~\ref{sec:results} presents the results for the electronic branching
fraction, the width, and the invisible width of the $W$ boson. Finally, we
state our conclusions in Section~\ref{sec:conclusions}.    
More extensive descriptions of the methods used in
this analysis can be found in Refs.~\cite{jamalsthesis}
and~\cite{ggthesis}.  


\section{The \Dzero\ Detector}
\label{sec:detector}
The \Dzero\ detector, described in detail elsewhere~\cite{d0nim},
consists of four major components: a non-magnetic central tracking
system for measuring the trajectories of charged particles; hermetic central
and end uranium/liquid-argon sampling calorimeters for measuring the
energies of electrons, photons, and hadrons; a toroidal spectrometer
outside of the calorimeter used for measuring the momenta of muons;
and a set of scintillation counters mounted on the front face of the
forward calorimeters used to detect inelastic \ppbar\ collisions and
measure the luminosity. We use a coordinate system where $\theta$
and $\phi$ are the polar and azimuthal angles, respectively, 
relative to the proton beam direction $z$.  The pseudorapidity 
$\eta$ is defined as $-\ln(\tan\frac{\theta}{2})$, and $\rho$ is the 
perpendicular distance from the beam line. 



The portions of the central tracking system used in this analysis
consist of four detector subsystems: a vertex drift chamber (\VTX),
a central drift chamber (\CDC) covering the pseudorapidity region
\abseta\ $<$ 1.1, and two forward drift chambers (\FDC) covering 1.1
$<$ \abseta\ $<$ 3.5.  The central tracking system provides a
measurement of the energy loss due to ionization (\dedx) for tracks 
within their tracking volume. This information can be used to help
distinguish between prompt electrons from $W$ and $Z$ boson decays and \ee\ 
pairs due to photon conversions. 

The calorimeter consists of three parts, a central calorimeter (\CC)
and two end calorimeters (\EC).  The calorimeters are segmented
longitudinally into an inner electromagnetic section (\EM) and an
outer hadronic section (\HAD).  The \EM\ calorimeter is 
segmented longitudinally into four layers, the third being at
the shower maximum for electromagnetic showers.
The calorimeter is segmented transversely in
towers, each covering approximately $\delta \eta \times \delta \phi =
0.1 \times 0.1$, with a further segmentation of $0.05 \times 0.05$ in
the third \EM\ layer.  
The third layer of the \CC\ is located at $\rho = 91.6$ cm, that of the
\EC\ calorimeter is located at $z = 178.9$ cm.
The \CC\ electromagnetic calorimeter
covers \abseta\ $\leq$ 1.1, while the \EC\ electromagnetic calorimeter
covers 1.4 $\leq$ \abseta\ $\leq$ 4.2.  The hadronic calorimeter
system provides full coverage to \abseta\ $ \leq 4.2$.  

The scintillation counters (\lzero) used for measuring luminosity 
consist of two layers of 1.6 cm thick scintillators covering
1.9 $\leq$ \abseta\ $\leq$ 4.3. Each layer has ten short (7 cm $\times$ 7 cm) 
scintillators, each glued to a single photomultiplier tube (PMT), and four 
long (7 cm $\times$ 65 cm) scintillators, each glued to two PMTs, one at each 
end. The average time resolution is 240 ps for the short scintillators and 
510 ps for the long ones. The two layers are oriented perpendicular to one 
another. The counters are located at $z = \pm 140$ cm on the front faces of 
the EC calorimeters, and provide a fast interaction trigger (within 800 ns) 
and a vertex resolution of 15 cm. 


\section{Data Selection}
\label{sec:dataselection}
\subsection{Event Topology}

Candidate \zb\ and \wb\ boson events are identified through their decay to two
electrons\footnote{Henceforth, the term ``electron'' refers generically to 
electrons and positrons.} which have an invariant mass consistent with the
mass of the \zb\ boson, or to an electron and a neutrino, respectively.
Electrons from \wb\ and \zb\ boson decays typically have large transverse 
energy \et\ and are isolated from other particles. They
are associated with a track in the tracking system and with a large
deposit of energy in one of the \EM\ calorimeters.  Neutrinos do not interact
in the detector, and thus create apparent energy imbalance in an event.  For 
each \wb\ boson candidate event, we measure the energy imbalance 
in the plane transverse to the beam direction (\met), and attribute this to
the neutrino. 

The particles that balance the component of the \wb\ or \zb\ boson
momentum transverse to the beam axis are referred to as the
``recoil.'' Particles from the break-up of the proton and anti-proton
in the inelastic collision are referred to as the ``underlying event.''
Particles from the recoil and underlying event are
indistinguishable.  While in principle there should be
no net \et\ in the underlying event, effects of finite resolution can cause 
the measured vector sum of the \et\ values of the particles from
the underlying event to be nonzero, and the underlying event 
therefore contributes to the \et\ resolution of the recoil.
The neutrino \et\ corresponds to the negative of
the vector sum of the electron \et, the recoil \et, and the
\et\ of the underlying event.

\subsection{Trigger}

A three-level trigger system is employed to select $W$ and
$Z$ boson candidates.  At Level-0, the \zee\ and \wev\
triggers require the detection of an inelastic collision via
simultaneous hits in the 
forward and backward \lzero\ scintillation detectors.  The $z$
position of the interaction point is calculated using the relative timing of
the hits in the counters and is required to satisfy \absz\ $\lt$
97 cm.

Level-1 consists of a hardware trigger that sums calorimetric
energy in towers of size $\Delta \eta \times \Delta \phi = 0.2 \times
0.2$.  The \wev\ trigger requires that at least one such \EM\ tower contain
transverse energy above a threshold of 10 \GeV.  
The \zee\ trigger requires the presence of two \EM\ trigger towers with 
\et\ $>$ 7 \GeV.


At the last trigger stage, Level-2, the full detector information is read into
a system of computers.  Electrons are identified as isolated clusters
of energy in the \EM\ calorimeters with longitudinal and transverse
shower shapes consistent with those of electrons.  Neutrinos are
identified with the measured energy imbalance in the calorimeter in
the plane transverse to the beam axis.  At this stage, the
polar angles for calorimeter towers are calculated using the vertex
position determined by the \lzero\ counters.  The \wev\ trigger
requires an electron candidate with \et\ $>$ 20 \GeV\ and \met\ $>$ 15
\GeV.  The \zee\ trigger requires two electron candidates with \et\
$>$ 20 \GeV.
Events passing the \wev\ or \zee\ triggers are written to magnetic
tape for subsequent analysis.

Additional requirements to ensure a well understood calorimeter response
and to cancel luminosity-dependent effects result in some data loss.
The Main Ring component of the Tevatron accelerator system passes
through the outer part of the hadronic calorimeter.  Beam losses from
the Main Ring can create significant energy deposits in the calorimeter,
resulting in large false \met.  The largest losses occur
when beam is being injected into the Main Ring. Events occurring
within a 400 ms window of injection are rejected, leading to only
a small loss of data.  Large beam losses can also occur when particles in the 
Main Ring pass through the \Dzero\ detector. Events within a 1.6 $\mu$s
window around these time periods are also rejected, resulting in an
approximately 8\% loss of data.
At the highest luminosities, the \wev\ 
trigger was prescaled by a factor of two to reduce the trigger rate to an
acceptable level. It was not necessary to
prescale the \zee\ trigger. To ensure that luminosity-dependent
effects cancel in the ratio of the cross sections, we discard runs
with a \wev\ prescale or with no \wev\ trigger, resulting in a loss of
approximately 32\% of the available \zee\ events.

\subsection{Offline Analysis Requirements}
\label{sec:analysis_requirements}

Offline, events passing the \wb\ or \zb\ trigger requirements
are studied for the presence of high-\et, isolated electrons and high \met\ 
which indicate the production and decay of \wb\ or \zb\ bosons. 
Electrons are required to have transverse and longitudinal shower
shapes consistent with those observed in test beam
studies~\cite{testbeam}.  In addition, they are
required to be isolated from
other calorimetric energy deposits and to have at least 95\% 
of their energy in
the \EM\ section of the calorimeter. To be considered isolated, 
electrons must satisfy the isolation requirement
\begin{equation}
\rm{Iso} \equiv \frac{E_{0.4}-E_{0.2}^{EM}}{E_{0.2}^{EM}} < 0.15 ,
\end{equation}
where $E_{0.4}$ is the total energy in a cone of radius 
$R = \sqrt{\Delta \eta^2 + \Delta \phi^2} = 0.4$ around the electron direction,
and $E_{0.2}^{EM}$ is the energy in a cone of radius $R = 0.2$ around the 
electron direction summed over the electromagnetic calorimeter only.

Geometric, or fiducial, requirements on the electrons are imposed
to ensure a well understood response from the calorimeters.
The electron position is measured in the third layer of the \EM\ calorimeter, 
where the resolution is best due to fine segmentation. 
We require the pseudorapidity of the electron calculated with respect to the
center of the detector, $\eta_D$, to satisfy $|\eta_D| < 1.1$ or 
$1.5 < |\eta_D| < 2.5$. In addition, for electrons in the \CC\ 
($|\eta_D| < 1.1$), we require that they be at least $0.05 \times 2\pi / 32$
radians away from any of the 32 \EM\ calorimeter inner module boundaries, 
thereby removing 5\% of the cell volume at each boundary. 

Finally, electrons in \wev\ candidate events and at least one of the electrons
in \zee\ candidate events are required to have a matching track\footnote{A 
matching track is a track which satisfies the track match significance 
requirement defined below.} whose position
extrapolated into the calorimeter agrees with the \EM\ cluster
position. To increase the size of the \zee\ sample, only one of the
electron candidates is required to have a matching track; electrons without a 
matching track are called ``loose''
electrons, while those with a matching track are called ``tight'' electrons.
The track match significance, $S_{\rm trk}$, is defined in terms of the 
distance between the extrapolated track and the \EM\ cluster centroid, and
the resolution in the distance: 
\begin{equation}
  S_{\rm trk} = \sqrt{
  { {(\rho\Delta\phi)^2} \over {\delta^2_{\rho\phi}} }+
  { {\Delta z^2} \over {\delta_z^2} }
  } 
  \,\,\,\,\, \rm{and} \,\,\,\,\,
  S_{\rm trk} = \sqrt{
  { {(\rho\Delta\phi)^2} \over {\delta^2_{\rho\phi}} }+
  { {\Delta \rho^2} \over {\delta_{\rho}^2} }
  }
\label{eq:trkmatch}
\end{equation}
for \CC\ and \EC\ electrons, respectively. Here, $\rho\Delta\phi$, $\Delta z$,
and $\Delta\rho$ are the distances in the azimuthal direction, the $z$ 
direction, and the radial direction respectively, and $\delta_{\rho\phi}$, 
$\delta_z$ and $\delta_{\rho}$ are the corresponding resolutions.
The longitudinal and transverse resolutions are $\delta_{z} = 1.7$ cm and
$\delta_{\rho\phi} = 0.3$ cm, respectively, in the CC, and 
$\delta_{\rho} =  0.7$ cm, $\delta_{\rho\phi} = 0.3$ cm in the EC.
The track match significance is
required to be less than 5 for candidates with \absetad $\lt$ 1.1, 
and less than 10 for candidates with 1.5 $\lt$ \absetad $\lt$ 2.5. 
Electron energies are corrected using the  electromagnetic energy scale
measured in the test beam, and adjusted to make the peak of the \zee\ 
invariant mass agree with the known mass~\cite{lepzmass} of the \zb\ boson. 
The electron energy scale is described in detail in Ref.~\cite{wmassprd}.

\begin{figure}[h!]
\center
\centerline{\psfig{figure=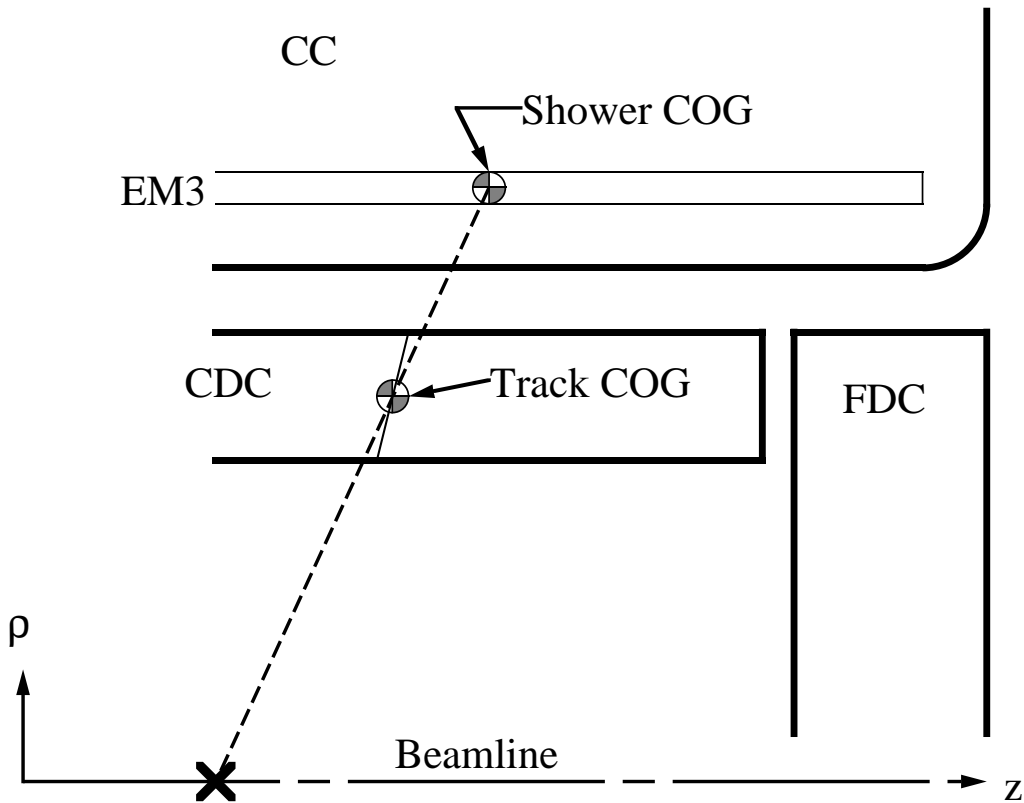,width=\figsize}}
        \caption{
The vertex position calculated using the position of the
electron cluster (as determined using the information from the third layer
of the \EM\ calorimeter) and the center-of-gravity of the electron track (as
measured in the tracking chambers).
}  \label{fig:vertexdef}
\end{figure}
\begin{figure}[h!]
\center
\centerline{\psfig{figure=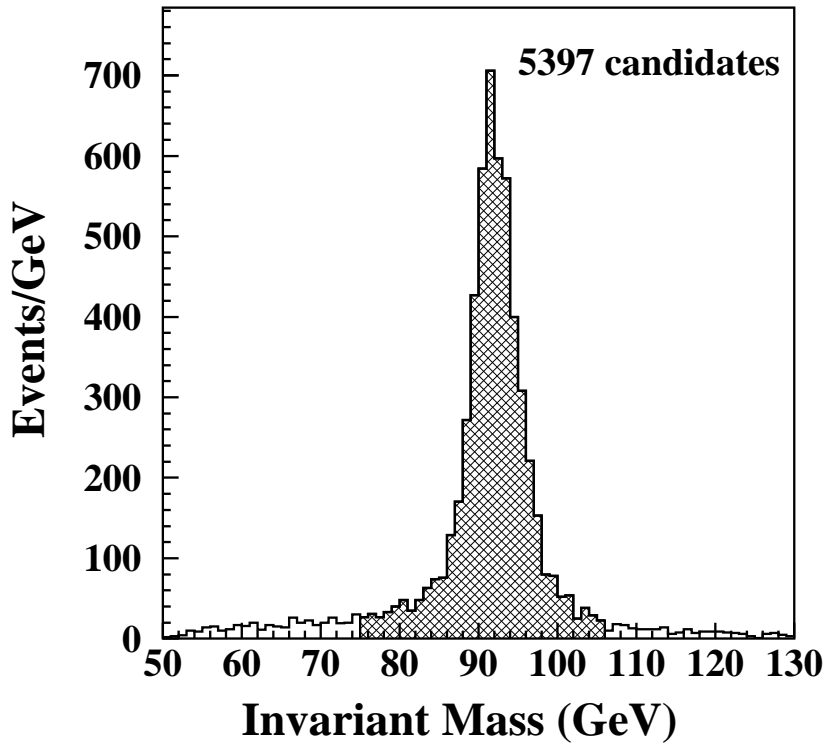,height=\figsize}}
        \caption{
The invariant mass distribution from the \zee\ candidate
event sample. The shaded region represents the dielectron invariant mass
requirement.
}  \label{fig:zmass}
\end{figure}

Candidates for the process \zee\ are required to have two electrons with 
\et\ $>$ 25 \GeV. The invariant mass of the dielectron pair is required to
satisfy 75 $<$ \mee\ $<$ 105 \GeV.  
The $z$ position of the event vertex is defined by the line connecting the 
center of gravity (COG) calorimeter position of the tight electron with the 
smallest $|\eta_D|$ and the COG position of its associated track, extrapolated 
to the beamline, as shown pictorially in Fig.~\ref{fig:vertexdef}.
The cluster COG position is calculated in the third, finely segmented, layer 
of the calorimeter. 
The track position is extracted at a $\rho$ of 62.0 cm for \CDC\ tracks or at 
a $z$ of 105.5 cm for \FDC\ tracks. The interaction vertex defined this way is
called the ``electron'' vertex and is required to be within \absz\ $\lt$ 97 cm.
A total of 5397 events passes the \zee\ selection criteria, of which 2737 
events have both electrons in the \CC\ calorimeter (\CC-\CC\ events), 2142 
events have one in the \CC\ and one in the \EC\ (\CC-\EC\ events), and 518 
events have both electrons in the \EC\ calorimeter (\EC-\EC\ events). 
Figure~\ref{fig:zmass} shows the invariant mass distribution of the \zee\ 
candidates.

\begin{figure}[h!]
\center
\centerline{\psfig{figure=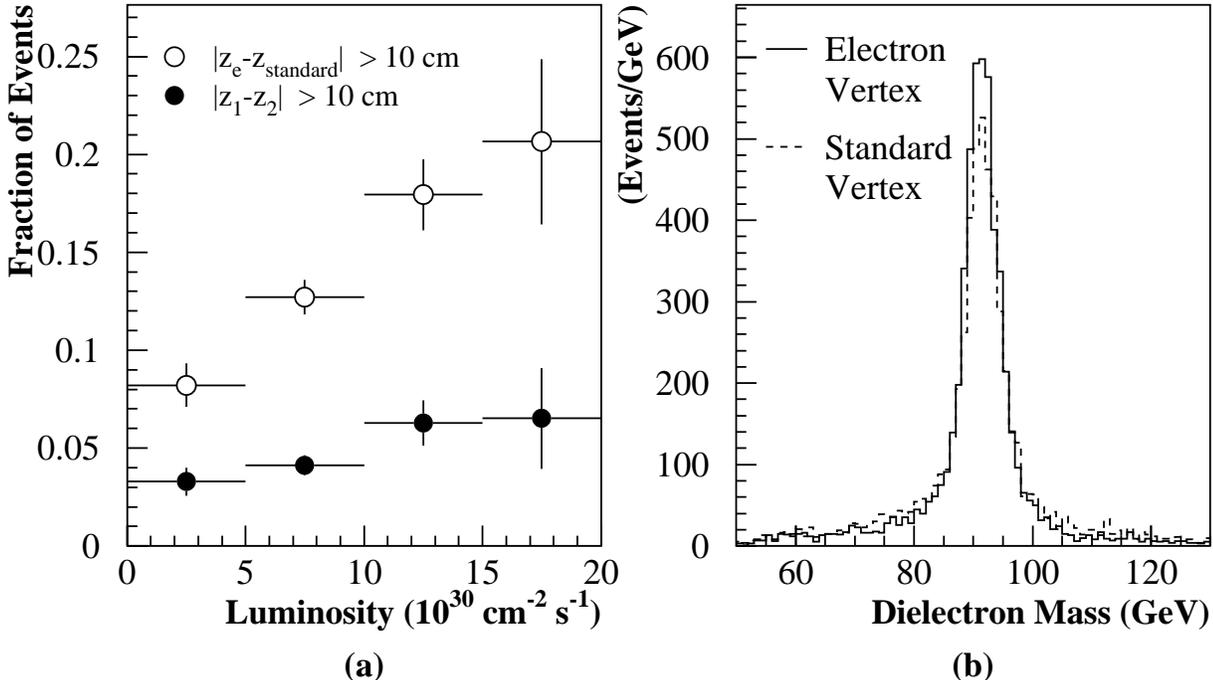,width=180mm}}
        \caption{
(a) Frequency at which the standard vertex, $z_{\rm{standard}}$ 
(calculated using all tracks), is more than 10 cm
away from the extrapolated track position, $z_e$, as a function of
luminosity.  For comparison, the frequency at which
\zee\ events with two tight electrons have extrapolated track
positions $z_1$ and $z_2$ differing by more than 10 cm is shown.
(b) Invariant mass distribution for \zee\ events when the
standard vertex position is used and when the electron
vertex position is used.
}  \label{fig:vertmisreco}
\end{figure}

\begin{figure}[hp!]
\center
\centerline{\psfig{figure=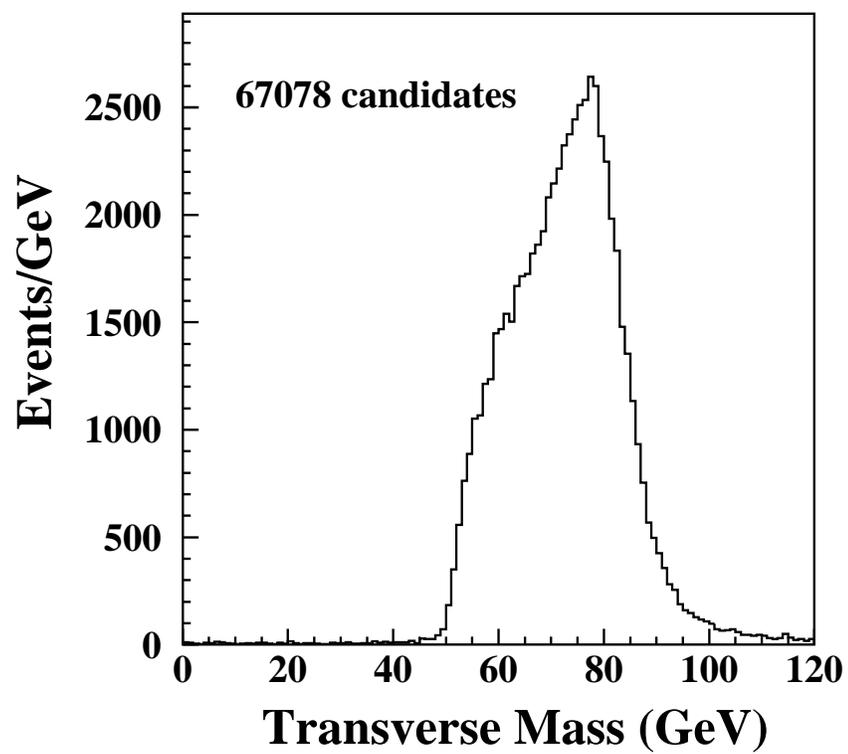,height=\figsize}}
        \caption{ The transverse mass distribution from the
\wev\ candidate event sample.	
}  \label{fig:wmass}
\end{figure}

Candidates for the process \wev\ are required to have one tight electron 
with \et\ $>$ 25 \GeV, and \met\ $>$ 25 \GeV.
Events containing a second loose or tight electron with \et\ $>$ 25 \GeV\ are
rejected to reduce backgrounds from $Z/\gamma^{*}\rightarrow ee$ events. 
The \met\ is calculated as the 
negative of the vector sum of the electron \et\ and the underlying event and 
recoil \et. The \et\ from the underlying event and the recoil is calculated as
the vector sum of the \et\ of all calorimeter cells except those
which contain the electron. 
While the electron \et\ is calculated using the above vertex, the
underlying event and recoil \et\ are calculated using a vertex
determined from all tracks in the \CDC, called the ``standard'' 
vertex, since the electron vertex is not available at the appropriate stage 
of the event reconstruction. The use of different vertex definitions results 
in a small degradation of the \met\ resolution. 
For the collected data, the mean number of interactions per crossing is
approximately 1.6.
For events with more than one interaction vertex, the one with
the largest number of associated tracks is selected as the standard vertex 
(even though it may or may not be the vertex closest to the extrapolated
position of the electron track). Figure~\ref{fig:vertmisreco} shows
the fraction of events in which the standard vertex is more than
10 cm away from the electron vertex. The figure
also shows the \zee\ invariant mass distribution when the standard vertex is
used and when the electron vertex is used. The electron vertex
gives a sharper invariant mass distribution, because it has better resolution 
and little luminosity dependence. 
A total of 67078 events passes the  \wev\ requirements, of which 46792 events
have their electron in the \CC\ (\CC\ events), with 20286 events in the \EC\ 
(\EC\ events).
Figure~\ref{fig:wmass} shows the transverse mass distribution of the
candidates, where the transverse mass is calculated as 
\begin{equation}
M_T = \sqrt{2 E_T(e) \met (1 - \cos{\delta \phi})}
\end{equation}
and $\delta \phi$ is the angle
between the electron and the \met\ in the transverse plane.

\section{Acceptances and Correction Factors}
\label{sec:acceptance}
\subsection{Monte Carlo Simulation}

The geometric and kinematic acceptances of the selection criteria are 
calculated using a Monte Carlo simulation.
Initially, the $W$ or $Z$ boson, the recoil system, and the underlying event
are generated with appropriate kinematic properties and the \wb\ or 
\zb\ boson is forced to decay in the electron channel.  
A second stage then models the response of the detector
and the effect of the geometric and kinematic selection criteria.

The primary event generator, originally
developed for the \Dzero\ $W$ boson mass analysis, 
is described in detail in Refs.~\cite{ericthesis,ianthesis}.  
The detector simulation was re-tuned for this analysis, 
because the mass analysis used only electrons with \absetad\ $\lt$ 1.1 and  
imposed different fiducial cuts at the azimuthal
boundaries of the central calorimeter modules. 
Also, the mass analysis was restricted to
events with \wb\ boson \pt\ $< 15$ \GeV, while this is not the case for
the present analysis.
The mass distribution of the \wb\ or \zb\ boson is  generated according 
to a Breit-Wigner distribution
convoluted with the CTEQ4M~\cite{CTEQ} parton distribution
functions, taking account of polarization in the decay.
The transverse momentum and
rapidity distributions of the \wb\ or \zb\ boson are generated by computing the
differential cross section, $d^2\sigma / dp_{T}^{2}dy$, using a 
program provided 
by Ladinsky and Yuan~\cite{LY}, as discussed in Ref.~\cite{wmassprd}. 
The \wb\ or \zb\ boson decays include the effects of lowest-order
internal bremsstrahlung, where a photon is radiated from a
final state electron, using the Berends-Kleiss calculation~\cite{berends}.  
This calculation predicts that 
approximately 31\% of the \wb\ boson events and 66\% of the \zb\ boson events 
have a photon with an energy above 50 \MeV\ in the final state.
In the simulation, the energies of the photon and its associated
electron are combined if their separation, $\sqrt{\Delta \eta^2 +
\Delta \phi^2}$, is less than 0.3, where $\Delta\phi$ is in radians.  
For the \zb\ boson, events are
generated according to the \zb\ boson line shape, and no Drell-Yan or
interference terms are included. 
The generator produces \wb\ and \zb\ bosons only over a finite mass range, 
and we
include a small correction in the acceptance to account for this.  As
a cross check, we have also calculated the acceptances using events generated 
with the \PYTH~\cite{pythia} event generator, and the 
results are consistent with those from our generator.

In the detector modeling phase of the simulation, the primary vertex
distribution is generated as a Gaussian with a width of 27 cm and a mean
position of $-0.6$ cm, to match the observed distribution.  Electron
energies and angles are smeared according to measured resolutions and
are corrected for offsets in energy scale due to
contamination from particles from the underlying event
or the recoil in the calorimeter towers containing the electron signal.
The electron energy scale is adjusted to reproduce the known 
mass~\cite{lepzmass} of the \zb\ boson.
The electron energy and angular resolutions used in the Monte Carlo
are tuned to reproduce the observed width of the \zee\ invariant mass
distribution for the sample used in this analysis.  

The uncertainty in the electromagnetic energy scale is 0.1\% for the \CC\ and 
1.6\% for the \EC.
The large uncertainty in the \EC\ energy scale is due to a
rapidity dependent calibration inaccuracy of the
\EC\ calorimeter. We correct for it in each sample which contains
\EC\ electrons (\CC-\EC\ \zee\ events, \EC-\EC\ \zee\ 
events, and \EC\ \wev\ events), by fitting the corresponding invariant or 
transverse mass distributions to the data, 
and the uncertainty is taken as the size of the correction. 

\begin{figure}[hp!]
\center
\centerline{\psfig{figure=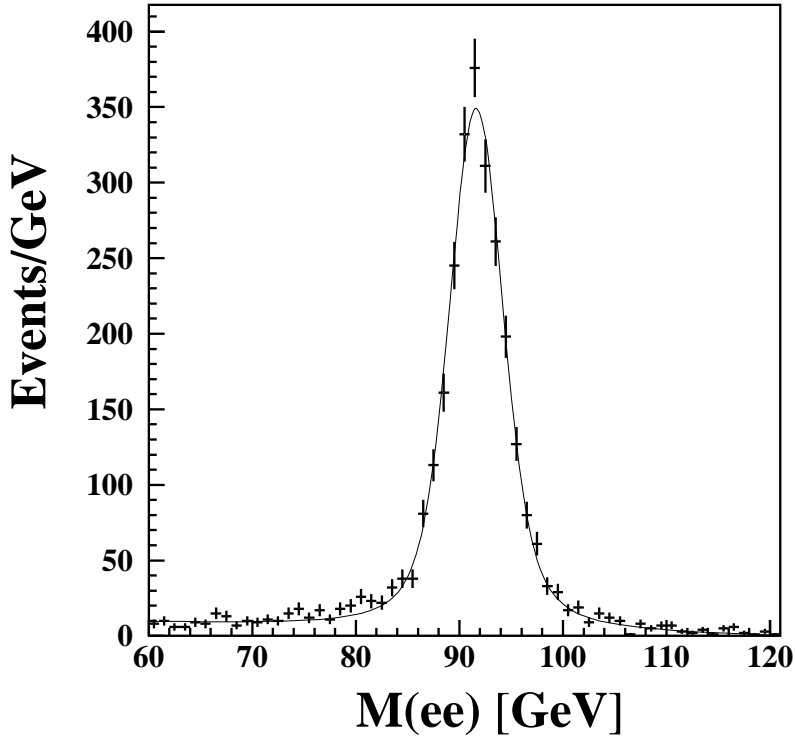,height=\figsize}}
        \caption{
Invariant mass distribution for the \CC-\CC\ \zee\ data sample. 
A Breit-Wigner convoluted with a Gaussian resolution is
fit to this distribution and the width is used to determine the constant
term in the \CC\ electron energy resolution. The $\chi^{2}$ per degree of 
freedom for the fit is 88.7/56.
}  \label{fig:constant_1}
\end{figure}

\begin{figure}[hp!]
\center
\centerline{\psfig{figure=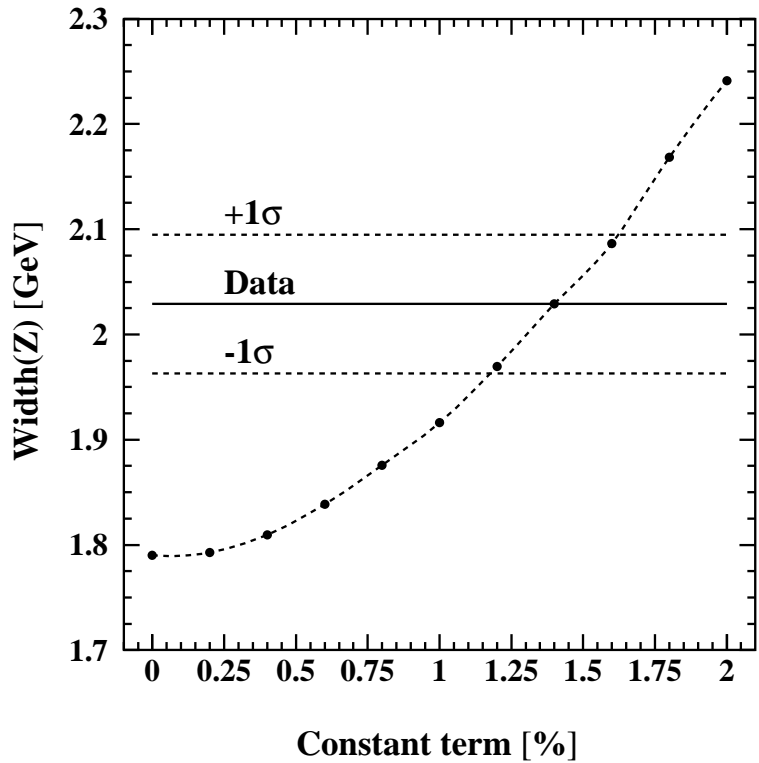,height=\figsize}}
        \caption{ 
Determination of the constant term for the electron energy resolution.
The curved dashed line connecting the Monte Carlo points shows the correlation
between the constant term in the \CC\ electron energy resolution and the 
fitted width of the \CC-\CC\ \zee\ invariant mass distribution from the Monte 
Carlo. The horizontal solid line shows
the fitted width of the \CC-\CC\ data sample, and the horizontal 
dashed lines the uncertainty on the fitted width. From the intersection
of the data line with the curved dashed line we determine the constant term 
for \CC\ electrons to be 0.014 $\pm$ 0.002.
}
\label{fig:constant_2}
\end{figure}

The electron energy resolution ($\Delta E$) can be parametrized as
$\Delta E/E = {\cal C} \oplus {\cal S} / \sqrt{E_{T}}$, where the two terms 
are called the constant and sampling term, respectively.
The value of $\cal S$ is known to high precision from test beam studies
and is $0.135\ \rm{GeV}^{1/2}$ for \CC\ electrons and $0.157\ \rm{GeV}^{1/2}$ 
for \EC\ electrons.  
The value of $\cal C$ in the simulation is adjusted until the r.m.s. from
the Monte Carlo \zee\ invariant mass distribution matches that of the data.
Figure~\ref{fig:constant_1} shows the result of fitting the 
invariant mass distribution of \CC-\CC\ \zee\ candidates
to a Breit-Wigner convoluted with a Gaussian.
Figure~\ref{fig:constant_2} shows the r.m.s. of the Gaussian that
is obtained when the same procedure is applied to Monte Carlo as a function
of the \CC\ constant term, along with the result from the data.  
The intersection of the two gives the constant term. 
The constant term in the \CC\ is thus determined to be 0.014 $\pm$ 0.002 with
the uncertainty being dominated by the statistics of the \zee\ sample.
The constant term in the \EC\ is $ 0.00 ^{+ 0.01}_{ - 0.00}$. 

The uncertainty in the polar angle of \CC\ electrons
is parametrized as an uncertainty in the
position of the track at a radius of 62 cm for \CDC\ tracks.
The $z$ position of the track at this radius has a 0.3 cm uncertainty.
The uncertainty in the polar angle for  \EC\ electrons
is absorbed into the large uncertainty in the \EC\ energy scale.

\begin{figure}[hp!]
\center
\centerline{\psfig{figure=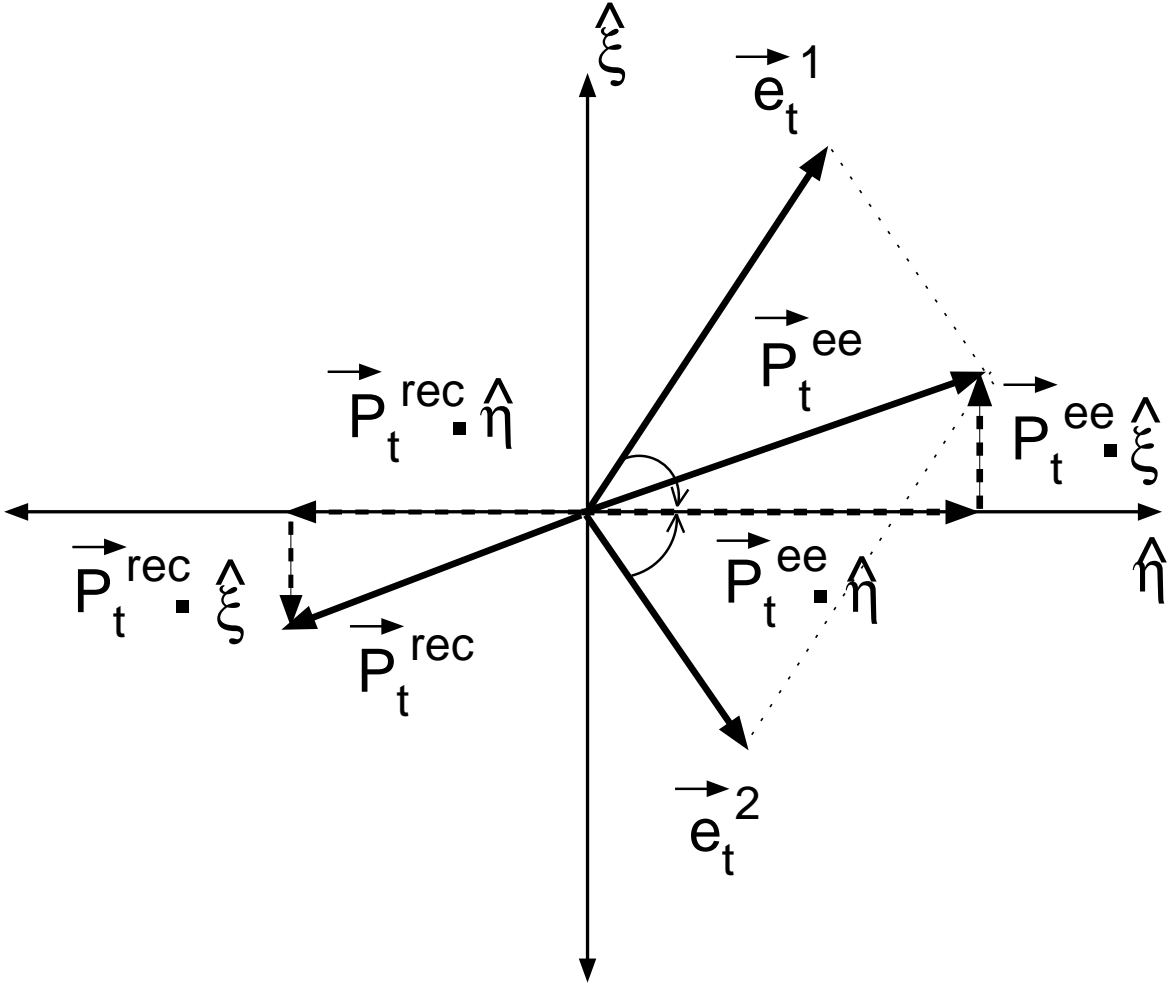,height=\figsize}}
        \caption{Definition of the $\hat{\eta}$--$\hat{\xi}$ coordinate system
in a \zee\ event. The $\hat{\eta}$ axis is the bisector of the electron 
directions in the transverse plane; the $\hat{\xi}$ axis is perpendicular to 
$\hat{\eta}$.
}
\label{fig:eta-xi}
\end{figure}

In the simulation, the recoil momentum is smeared by the measured 
resolution. The recoil is also corrected for any losses of particles to the 
same calorimeter towers as the electron. The model of the response of the 
calorimeter to particles recoiling against the \wb\ or \zb\ boson is tuned
using \zee\ events. The $\hat{\eta}$ axis is defined as the bisector of the 
azimuthal angle between the two electrons, as shown in Fig.~\ref{fig:eta-xi}.
We compare the component of the $p_T$ of the $Z$ boson along $\hat{\eta}$
as calculated using the energies of the electrons, $(p_T^{ee})_{\hat{\eta}}$,
to that calculated by summing the
transverse momentum of all towers in the calorimeter, except those
containing the electrons, $(p_T^{\rm{rec}})_{\hat{\eta}}$.

\begin{figure}[hp!]
\center
\centerline{\psfig{figure=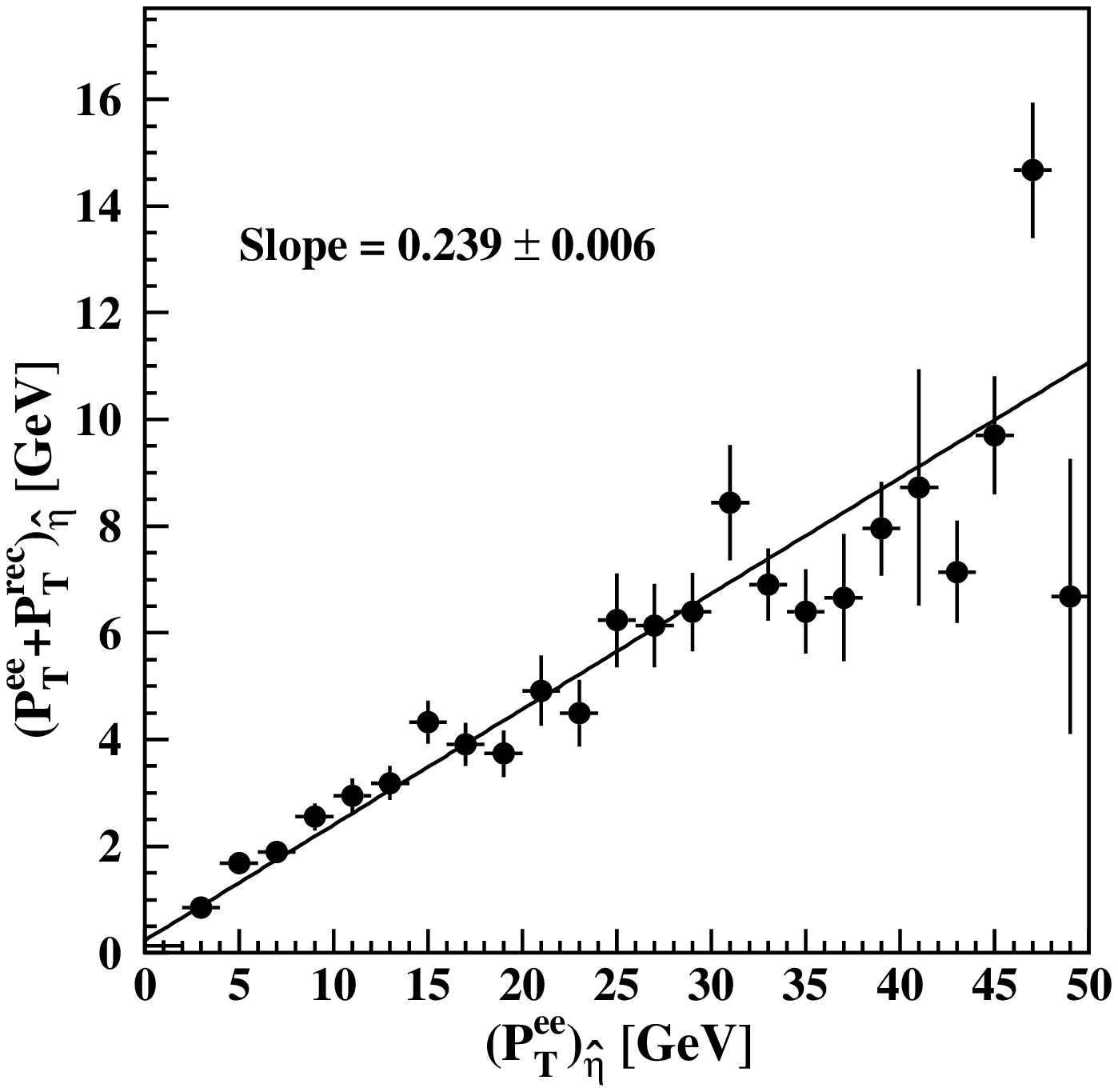,height=\figsize}}
        \caption{ 
The $\hat{\eta}$-imbalance, $(p_T^{\rm{rec}})_{\hat{\eta}} + (p_T^{\rm{ee}})_{\hat{\eta}}$, versus $(p_T^{ee})_{\hat{\eta}}$ from the \zee\ 
sample. The solid line is a linear 
fit to the data points, with a slope of 0.239 $\pm$ 0.006 and a $\chi^{2}$ per
degree of freedom of 47.5/23. Up to a $p_T$ of 25 \GeV, where most of the
\wev\ and \zee\ data is, the $\chi^{2}$ per degree of freedom is 1.2. The 
hadronic response contributes only a small fraction of the uncertainty in the 
acceptance.
}  \label{fig:had_tune_1}
\end{figure}

\begin{figure}[hp!]
\center
\centerline{\psfig{figure=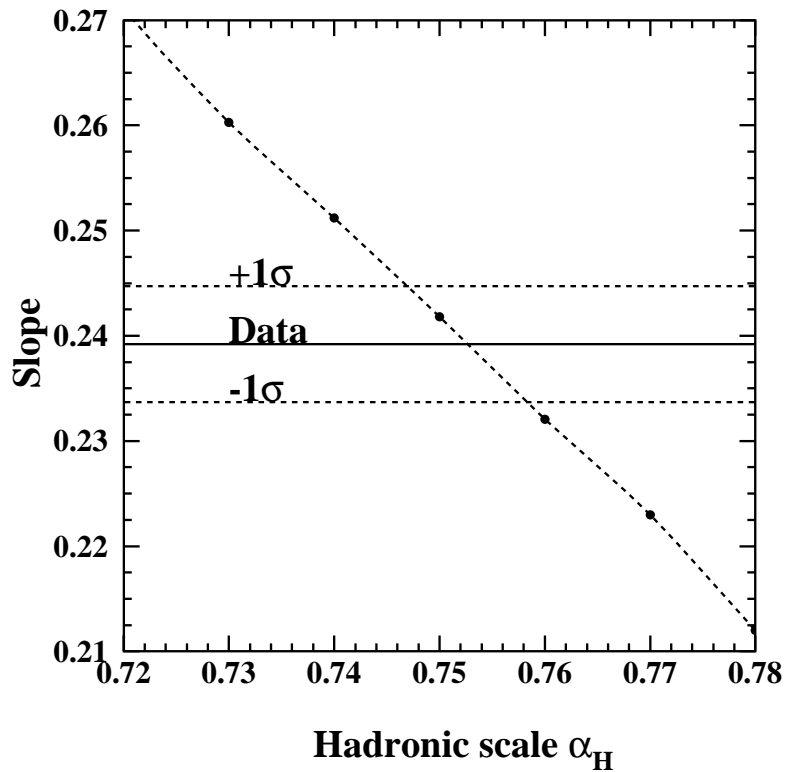,height=\figsize}}
        \caption{
Determination of the hadronic scale $\alpha_H$. The points represent the slope
of the line $(p_T^{\rm{rec}})_{\hat{\eta}} + (p_T^{\rm{ee}})_{\hat{\eta}}$ 
versus $(p_T^{ee})_{\hat{\eta}}$ obtained from 
Monte Carlo as a function of $\alpha_H$. 
The intersection of the dashed line connecting the Monte Carlo points 
with the solid line, obtained from data, determines the hadronic scale used 
in the simulation. We take $\alpha_H = 0.753 \pm 0.024$.
}  \label{fig:had_tune_2}
\end{figure}

Because the calorimeter response is different for electrons and for recoil
particles, the algebraic sum of $(p_T^{\rm{rec}})_{\hat{\eta}}$ and 
$(p_T^{\rm{ee}})_{\hat{\eta}}$ is on average non zero.
The average value of this ``$\hat{\eta}$-imbalance'' scales linearly with
$(p_T^{ee})_{\hat{\eta}}$, as shown in Fig.~\ref{fig:had_tune_1}.
The recoil scale used in the simulation is tuned such that
applying the same procedure to Monte Carlo events yields 
the same response as the data. Figure~\ref{fig:had_tune_2} 
shows the slope of the average 
$(p_T^{\rm{rec}})_{\hat{\eta}} + (p_T^{\rm{ee}})_{\hat{\eta}}$ versus 
$(p_T^{ee})_{\hat{\eta}}$ from the Monte Carlo as a function of the hadronic 
scale, along with the slope determined from data.  
The intersection of the two determines the hadronic response to be
$\alpha_H = 0.753 \pm 0.024$ relative to the electromagnetic energy scale, with
the uncertainty being dominated by uncertainties in the 
\EC\ electromagnetic energy scale. 
The hadronic energy resolution is parametrized in the same way as the electron
energy resolution, and is known from jet studies to have a constant term of
4\% and a sampling term of $0.8 / \sqrt{p_T/\rm{GeV}}$.  

\begin{figure}[hp!]
\center
\centerline{\psfig{figure=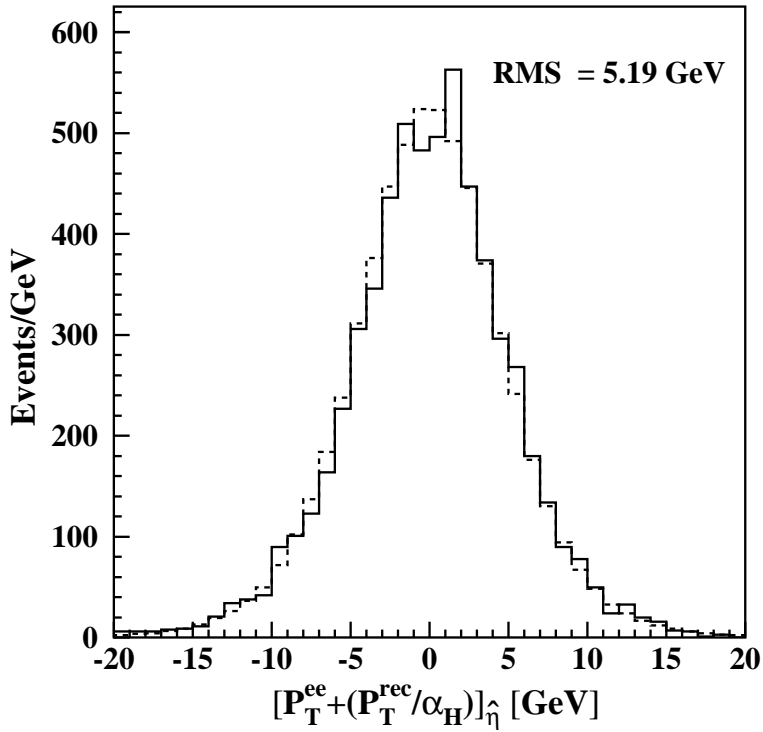,height=\figsize}}
        \caption{
Distribution of the ``$\hat{\eta}$-balance,'' the magnitude of the vectorial 
sum of $(p_T^{\rm{rec}})_{\hat{\eta}}/\alpha_{H}$ and 
$(p_T^{ee})_{\hat{\eta}}$, for events in the \zee\ sample (solid histogram), 
and for Monte Carlo (dashed histogram).
}  
\label{fig:had_tune_3}
\end{figure}

\begin{figure}[hp!]
\center
\centerline{\psfig{figure=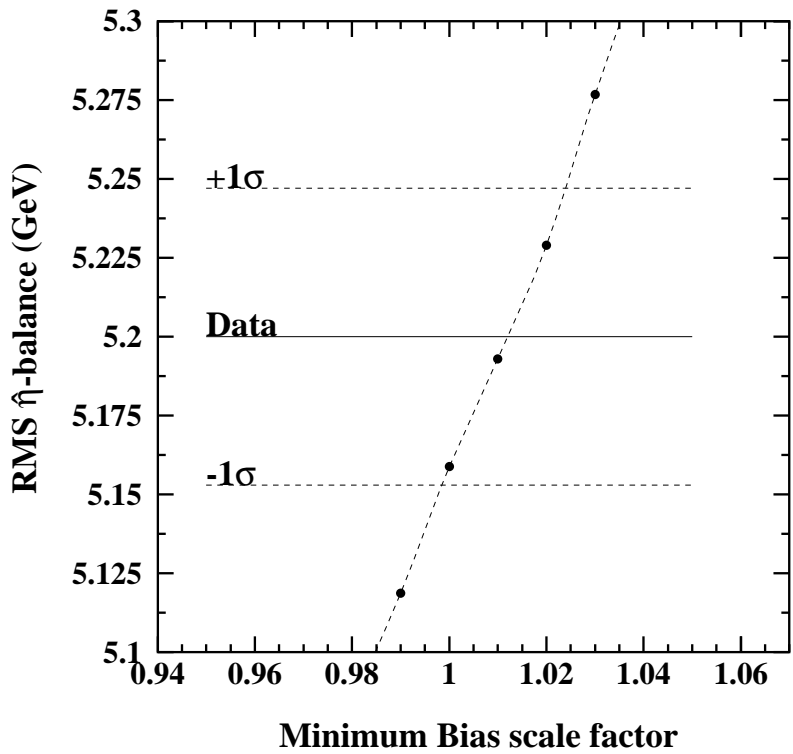,height=\figsize}}
        \caption{
Determination of the minimum bias scale factor.
The points represent the r.m.s. of the $\hat{\eta}$-balance distribution
from Monte Carlo as a function of the minimum bias scale factor.
The solid horizontal line shows the r.m.s. from the data sample.
The intersection of the dashed line connecting the Monte Carlo points 
with the data line determines the minimum bias scale factor used in the
simulation to be $1.01 \pm 0.02$.
}  
\label{fig:had_tune_4}
\end{figure}

The underlying event is modeled using events taken with a
\lzero\ trigger (minimum bias events) with the same luminosity
profile as the \wb\ and \zb\ boson samples.
We pick a minimum bias event randomly from this sample 
and its \met\ is combined vectorially with that of the simulated \wb\ boson. 
To account for any possible difference between the underlying event in 
\wb\ boson and in minimum bias events, we introduce a multiplicative scale 
factor for the \et\ of the minimum bias events.
The scale factor is estimated using the \zee\ sample and set so that the width
of the ``$\hat{\eta}$-balance'' distribution from the simulation 
agrees with that from the data, where ``$\hat{\eta}$-balance'' is  
$(p_T^{\rm{rec}}/\alpha_H + p_T^{ee})_{\hat{\eta}}$.
Figure~\ref{fig:had_tune_3} shows this quantity
for the \zee\ event sample. Figure~\ref{fig:had_tune_4} shows 
the r.m.s. of the $(p_T^{\rm{rec}})_{\hat{\eta}}$ distribution from the 
simulation as a function of the minimum bias scale factor.  
The simulation has the same r.m.s. as the data when
the scale factor between the minimum bias events and the $W$ boson underlying 
events is 1.01 $\pm$ 0.02.

\begin{figure}[hp!]
\center
\centerline{\psfig{figure=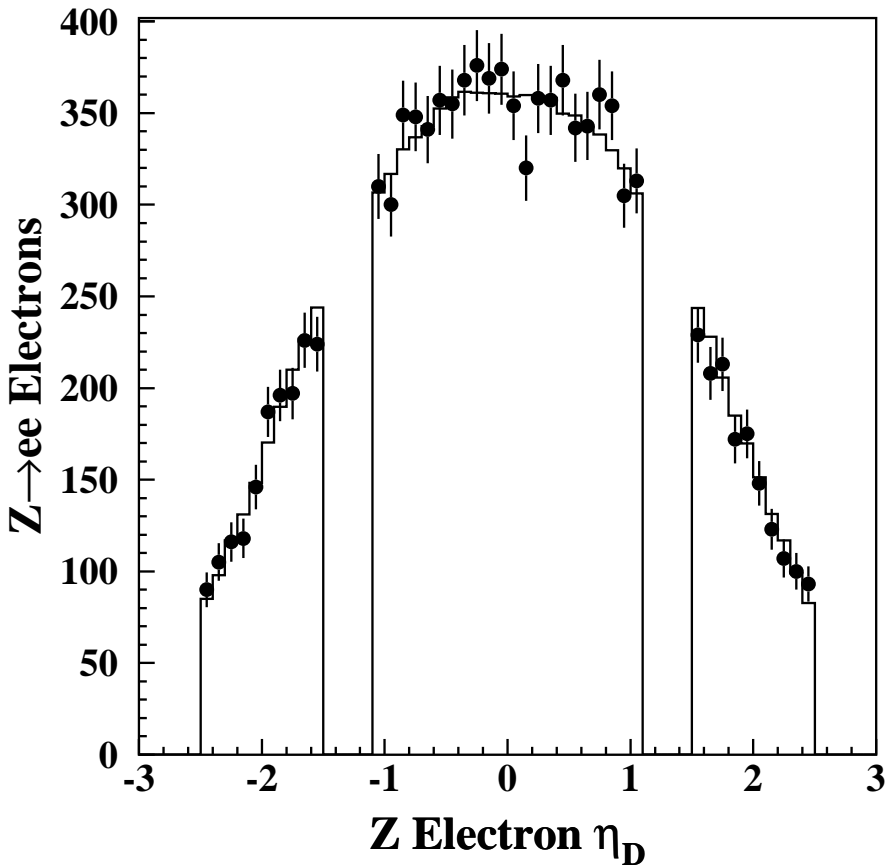,height=\figsize}}
        \caption{
The electron $\eta_D$ distribution for \zee\ candidates (solid circles) and for
the Monte Carlo (histogram) after all corrections and cuts except for track
match have been applied. 
The error bars represent the statistical uncertainty in the data. The $\chi^2$
per degree of freedom is 0.65.
}  
\label{fig:deteta}
\end{figure}

Figure~\ref{fig:deteta} shows the electron detector pseudorapidity 
distribution, 
$\eta_D$, for \zee\ candidates and for the Monte Carlo after all cuts and 
corrections have been applied. The sharp edges 
correspond to the fiducial requirements applied to the electrons. The data
and the Monte Carlo agree well. Since the tracking efficiency is obtained from
the \zb\ data (as explained in Sec.~\ref{sec:efficiencies}), the figure shows
electrons without the tracking requirement. 

\subsection{Geometric and Kinematic Acceptance}

The acceptance is defined as the fraction of generated \wev\ or \zee\ events
satisfying the kinematic and geometric requirements.
Samples of 25,000,000 events are used to estimate all systematic uncertainties,
except  those from ambiguities in the parton distribution functions and 
differences in generators.
For these, we use the slower \PYTH\ generator and samples of 
1,000,000 events,
corresponding to statistical errors of 0.1\%, which are small compared to the
dominant uncertainties. Table~\ref{tab:accep-syst} shows the acceptance results
and a summary of the systematic uncertainties.

\begin{table}[ht!]
\begin{center}
\caption{
Acceptances and their systematic uncertainties
for $W$ and $Z$ boson events and their ratio. }
\medskip
\begin{tabular}{|l|c|c|c|}
\hline 
  & $A_W$ & $A_Z$ & $\frac{A_W}{A_Z}$ \\
 & & & \\
\hline
 Acceptance & \wacc\ $\pm$ \waccerr & \zacc\ $\pm$ \zaccerr & 
              \racc\ $\pm$ \raccerr \\
 & & & \\
\hline \hline
 {Error source} &
  {$ \frac{ \delta A_W }{ A_W } $ [\%]} &
  {$ \frac{ \delta A_Z }{ A_Z } $ [\%]} &
  {$ \delta \left( \frac{ A_Z }{ A_W } \right) /
                 \left( \frac{ A_Z }{ A_W } \right) $ [\%]}  \\ 
& & & \\
\hline
\pt\ spectrum           & 0.096   & 0.104   & 0.100 \\
Parton distribution functions     & 0.189   & 0.314   & 0.252 \\
Clustering algorithm      & 0.141   & 0.294   & 0.153 \\
$\delta M_W$            & 0.130   & ---     & 0.130 \\
$\delta \Gamma_W$       & 0.050    & ---    & 0.050 \\
EM energy scale         & 0.685   & 0.337   & 0.698 \\
EM energy resolution    & 0.024   & 0.037   & 0.044 \\
Hadronic response       & 0.129   & ---     & 0.129 \\
Hadronic resolution     & 0.078   & ---     & 0.078 \\
Angular resolution      & 0.019   & 0.046   & 0.027 \\ 
Generated mass  range         & 0.150   & 0.180   & 0.234 \\
Generator     & 0.343   & 0.516   & 0.172 \\
\hline
Total       &            0.85\% & 0.78\% & 0.85\% \\ 
\hline 
\end{tabular}
\label{tab:accep-syst}
\end{center}
\end{table}

The uncertainties in the \wb\ and \zb\ boson \pt\ spectra are calculated by 
varying the theoretical parameters in Ref.~\cite{LY} within the range
quoted by the authors.
The systematic uncertainties from the choice of parton distribution
functions are calculated from the largest excursion in
acceptance found using the CTEQ4M~\cite{CTEQ}, CTEQ2M~\cite{CTEQ}, 
MRSD$-$~\cite{MRSD-}, MRS(G)~\cite{MRSG}, GRV94HO~\cite{GRV94HO},
and versions of the MRSA$'$ distribution functions with 
values of the strong coupling constant ranging from 0.150 
to 0.344~\cite{MRSAp}.

The systematic uncertainties in the acceptance due to the presence
of radiative photons in the event come from uncertainties on the minimum 
separation in $\eta$-$\phi$\ space the electron and the photon must have in 
order to be resolved as separate clusters by our calorimeter clustering 
algorithm. The uncertainties due to effects of the clustering algorithm
are calculated by varying the size 
of the cone that is used to decide
whether or not the photon will be resolved from the electron in the
detector between 0.2 and 0.4. 

We use a \wb\ boson mass of 80.375 \GeV\  and width of 2.066 \GeV, 
and vary these by $\pm 0.065$~\GeV\ and $\pm 0.060$~\GeV, respectively. 
The $W$ boson mass is the result of combining the measurements from 
\Dzero~\cite{wmassprd}, CDF~\cite{wmassCDF}, and a fit to all direct 
$W$ boson mass measurements from LEP~\cite{lepzmass}. The $W$ boson width is 
the current world average~\cite{Wpapers,ua1r,ua2r,cdf1aR,d01aR,LEPwwidth}.  
The systematic uncertainties in the acceptance due to the \EM\ energy
scale, \EM\ resolution, hadronic response, hadronic resolution, and
the resolution on the polar angle of electron tracks are found by
varying the relevant parameters within the Monte Carlo simulation by
their individual uncertainties. 

The generation of \wb\ and \zb\ bosons is limited to the mass ranges
40--120 \GeV\ for $W$ bosons and 30--150 \GeV\ for $Z$ bosons. The error
quoted on generated mass in Table~\ref{tab:accep-syst} is
the uncertainty on the fraction of events outside this mass
window that would pass our selection criteria.  The error is dominated
by the statistics of the Monte Carlo samples, but is well below the dominant
uncertainties.
The error quoted on the generator is from a comparison of the difference in 
acceptance between our Monte Carlo and \PYTH\ (after smearing 
\PYTH\ for detector response).

The acceptances and their uncertainties for \wev, \zee, and their ratio are 
shown in Table~\ref{tab:accep-syst}.
The \wb\ boson acceptance is $A_W = \wacc \pm \waccerr$. 
The largest contributions to the uncertainty arise from uncertainties in 
the \EM\ energy scale, the width of the \wb\ boson, 
the difference between our generator and \PYTH, 
and uncertainties in the parton distribution functions.
The acceptance for \zee\ events is $A_Z = \zacc \pm \zaccerr$. 
The largest sources of systematic uncertainty arise from the difference between
our generator and \PYTH, effects of the electron-photon 
clustering algorithm in radiative \zb\ boson decays, and uncertainties in the
\EM\ energy scale and in the parton distribution functions.
In the ratio of acceptances, a few of the systematic uncertainties
are reduced by partial cancelations of correlated errors. The ratio of the 
acceptances is  $A_Z/A_W = \racc \pm \raccerr$.

\subsection{Drell-Yan Correction}
\label{drellyancorr}

It is conventional to report \sigmaz\ as the product of the cross section and 
branching ratio, assuming the \zb\ boson as the only source of
dielectron events.  However, the production of dielectron events is
properly described by considering the \zb\ boson, the photon
propagator, and the interference between the two.  The Drell-Yan correction
factor relates the number of events in our mass window 
to what would be expected purely from \zb\ boson production.
To obtain this correction, we use \PYTH\ to generate events with
just the contribution from the $Z$ boson, and, separately, using the full 
Drell-Yan process with interference terms (combining \zb\ boson and photon 
diagrams).
We process both samples with the same Monte Carlo simulation
used for the acceptance calculation.  The ratio of the complete
Drell-Yan cross section ($\sigma_{DY}$) to the  cross
section for the \zb\ boson alone ($\sigma_{Z}$), for events passing our
\zee\ selection criteria, is estimated to be 
\begin{equation}
\sigma_{DY} / \sigma_Z  =
\frac{1}{1-f_{DY}} = 
1.012 \pm 0.001
\end{equation}
or $f_{DY}$ = \zfdy\ $\pm$ \zfdyerr\ 
as the fraction of production cross section attributable to the
presence of the photon propagator.  
The systematic uncertainty is evaluated by using the \ISAJ~\cite{isajet} 
generator instead of \PYTH\ and is estimated as the difference
between the two generators. 
The primary uncertainty in $f_{DY}$ is due to Monte Carlo statistics, but its 
contribution to the total uncertainty in the $Z$ boson cross section and in 
\R\ is negligible.

\subsection{NLO Electroweak Radiative Corrections}
Next to leading order (NLO) electroweak processes modify the cross
sections and their ratio~\cite{baer}. A full NLO calculation is available for
the \wb\ boson which suggests that the \wb\ boson cross section would decrease
by a multiplicative factor of $\nlow \pm \nlowerr$~\cite{baer_private}.  
For the \zb\ boson, only the full QED calculation is available; the 
purely weak part is missing.  
For the ratio $\cal R$, the best theoretical estimate at this time is a 
multiplicative factor of $\nlor \pm \nlorerr$~\cite{baer_private},  
where the uncertainty is dominated by the difference between the NLO 
corrections to the \wb\ and \zb\ boson cross sections, due mainly to the 
purely weak corrections missing in the $Z$ boson calculation. 
This theoretical uncertainty is expected to be reduced in the future. A 1\% 
uncertainty in $\cal R$ due to NLO electroweak radiative corrections is
quoted in this analysis. 



\section{Efficiencies}
\label{sec:efficiencies}
\subsection{Electron Identification Efficiencies}

Electron identification efficiencies are obtained using
\zee\ events selected by requiring two electron candidates satisfying
only standard kinematic and fiducial requirements. An electron
is considered a ``probe'' electron if the other electron
in the event passes all standard electron identification criteria. This gives
a clean and unbiased sample of electrons.
We count the number of events inside a \zb\ boson invariant mass $m(ee)$ window
before and after applying the electron identification criteria to {\it each} 
probe electron. The ratio of the number of events in the \zb\ boson mass 
window, after background subtraction, gives the electron identification 
efficiency.  
Two techniques are used to determine the background.  In the sideband
method, the number of \zee\ events in the regions 60 $<$ \mee\ $<$ 70 \GeV\ 
and 110 $<$ \mee\ $<$ 120 \GeV\ is used to estimate the number of
events inside the signal region by assuming a linear shape
for the background.
In the second method, backgrounds are estimated by fitting the
observed invariant mass distribution to a Breit-Wigner (smeared with a 
Gaussian to account for detector resolution) for the \zb\ boson, and an 
assumed first-order polynomial for the background.  The background is 
estimated from the contribution of the polynomial within the signal region.  
The difference between  these two estimates comprises a part of the systematic 
uncertainty, which also includes the sensitivity of the result to the band 
chosen as the signal region. 
We have also used an exponential
shape for the background, and the efficiencies resulting from such a fit are 
all well within the corresponding uncertainties.
We have checked for any dependence of the efficiency on the \et\ of the
electron, and find none. 

The above method does not yield the correct efficiency when the probability
for one of the electrons to pass the identification requirements is
correlated with that of the other.  We check for such
correlations in the calorimeter-based identification requirements 
using a \GEAN-based simulation~\cite{geant}.  We find
that  the impact of such a correlated bias
is small compared to
the uncertainty on the efficiency, and we neglect it.  For
tracking-based electron identification requirements, we
evaluate the correlations using data and find that these correlations
can not be neglected.  We select events with two electrons that pass the
geometric and calorimetric electron identification
requirements of the \zee\ data sample.  We count the background-subtracted
number of \zee\ events that have both electrons passing the tracking
requirements ($N_{PP}$), only one electron passing the tracking
requirements ($N_{PF}$), and with no electrons passing the
requirements ($N_{FF}$).  The efficiency for a \wb\ boson to pass the
tracking requirements is then $\left( 2 N_{PP} + N_{PF}
\right) / \left[ 2 \left( N_{PP} + N_{PF} + N_{FF} \right) \right] $. 
The efficiency
for a \zb\ boson to pass the tracking requirements is $\left( N_{PP} +
N_{PF} \right) / \left( N_{PP} + N_{PF} + N_{FF} \right)$.  The tracking 
efficiency for a \wb\ boson or a \zb\ boson  is found to be
1.7 $\pm$ 0.3\% lower than what one would get assuming no correlations. The
effect of this correlation cancels in the ratio of the cross 
sections~\cite{corr_example}.  

The efficiency of the calorimetric requirements is 
0.916 $\pm$ 0.006 for \CC\ electrons and 0.870 $\pm$ 0.007 for \EC\
electrons.  The efficiency of the tracking requirements is 0.777
$\pm$ 0.006 for \CC\ electrons  in $W$ boson events
and 0.731 $\pm$ 0.010 for \EC\ electrons in $W$ boson events. 
 (Because of the presence of correlations, the per track efficiency is 
not a useful concept for $Z$ boson events.)

\subsection{Trigger Efficiencies}

Trigger efficiencies are evaluated from different data samples.
A special trigger which is identical to the \wev\ trigger, except for
its \met\ requirement, is used to evaluate the relative efficiency of
the \met\ requirement in the $W$ boson trigger.  The efficiency 
of the \met\ requirement is found to be 0.993 $\pm$ 0.001.
The efficiency of the electron requirements in the trigger is
measured from a dielectron sample using the same method  
used  to determine the electron identification efficiencies,
and is found to be $0.995 \pm 0.001$ for electrons in the \CC, and
$0.996 \pm 0.002$ for electrons in the \EC.
A portion of the \wev\ data was taken without requiring the \lzero\
component of the trigger.  By studying these events, and taking into
account the luminosity-dependent effects, we find the \lzero\ efficiency for 
\wb\ boson events to be $0.986 \pm 0.005$. We assume that this efficiency is 
the same for \wev\ and for \zee\ events, and therefore cancels in the ratio.

\subsection{Total Efficiencies}

The efficiency for a \wb\ or \zb\ boson to pass the electron
identification requirements is obtained from
the convolution of the efficiencies with the acceptances as a function of
the $\eta$ of the electrons from our Monte Carlo. 
From our Monte Carlo simulation, of the events that pass our kinematic and 
geometric selection, 68.70\% of $W$ bosons have a \CC\ electron and 31.30\%
have an \EC\ electron.  For the \zb\ boson, 49.69\% have both
electrons in the \CC, 40.55\% have one \CC\ and
one \EC\  electron, and 9.76\% have both electrons in the \EC .
The  efficiency for \wb\ bosons to pass both the electron
identification and the trigger requirements is 0.685 $\pm$ 0.008.  The
analogous efficiency for \zb\ bosons is 0.754 $\pm$ 0.011.

Taking into account the \lzero\ efficiency, the total
efficiency  for \zb\ bosons
is $\epsilon_Z = \zeff \pm \zefferr$.  For the \wb\ boson,
combining  electron identification and trigger efficiencies with
the efficiencies of the \met\ and \lzero\ requirements, we obtain a
total efficiency of $\epsilon_W = \weff \pm \wefferr$.  The ratio of the
efficiencies is $\reff \pm \refferr$, where the error takes into
account the correlations between the uncertainties in the 
$W$ and $Z$ boson efficiencies.

\subsection{Diffractive Production of Weak Bosons}
Diffractive production of \wb\ and \zb\ bosons at the Tevatron occurs when 
the incident proton or anti-proton escapes intact, losing a small fraction of 
its initial forward momentum. 
Our cross section measurements include both diffractive and non-diffractive
$W$ and $Z$ boson production. The perturbative theoretical calculation of 
Ref.~\cite{theoryrs} does not include an explicit calculation of diffraction,
but diffraction contributions to the total cross sections enter through the
parton distribution functions.
A recent measurement~\cite{CDFdiffrw} 
reports the diffractive to non-diffractive $W$ boson production ratio to be 
$(1.15 \pm 0.55)\%$. No such measurement exists to date for $Z$ bosons, 
although it is believed that diffractive $Z$ boson production exists at 
roughly the same level. Recent theoretical calculations suggest that the ratio
of diffractive $W$ to $Z$ boson cross sections is roughly the same as the ratio
of inclusive\footnote{To obtain the inclusive cross section ratio, one needs 
to multiply \R\ times B(\zee)/B(\wev).} 
cross sections (see Table V of Ref.~\cite{thrydiffr}). 
Since the \lzero\ trigger requires simultaneous hits in the forward and 
backward scintillation counters, such events would not pass our selection 
unless accompanied by a minimum bias interaction.
The \lzero\ trigger efficiency is calculated from $W$ boson events without a 
\lzero\ requirement, and no correction is made to subtract diffractive $W$
bosons, so
in practice we account for all diffractive $W$ bosons produced. The same 
efficiency is used for $Z$ boson events under the assumption that the 
underlying events in $W$ and $Z$ boson production are essentially identical.
In order to have an appreciable effect on $R$, the diffractive production of 
$Z$ bosons would have to be several times larger than that observed for 
$W$ bosons, so we may safely neglect the effect on $R$.
The effects of diffractive production on the individual cross sections 
are much smaller than the luminosity uncertainty and are therefore neglected.



\section{Backgrounds}
\label{sec:backgrounds}
\subsection{Backgrounds from multijet, \mbox{\boldmath $b$} quark, and direct 
photon sources in the \mbox{\boldmath \wev\ } sample}

The fraction of background events in the \wev\ sample that is due to
multijet, $b$ quark, and direct photon sources, $f_{QCD}$ (also referred to 
as QCD background), is calculated by comparing the number of events in the 
\wev\ sample to the number in a sample with the same kinematic requirements,
but with loosened or tightened electron-identification requirements.
The larger sample is the ``parent'' sample, the smaller the
``child.''
If the efficiency for signal and for background to pass the
child criteria relative to the parent requirements is known, the
number of signal and background events can be simply calculated.  The
efficiency for electrons from \wb\ boson decay to pass the identification
requirements $(\epsilon_s)$ is calculated using the \zee\ sample.  The
efficiency for ``electrons'' from background sources $(\epsilon_b)$ is 
calculated using a data sample obtained using  the same criteria as the \wev\ 
sample, except 
requiring small \met\ in the event instead of large \met\ (to remove
$W$ boson events).  The main source of systematic uncertainty 
is from the assumption that the ``electrons" from background sources
in events with small \met\ have the same value for $\epsilon_b$ as
those with \met\ $>$ 25 \GeV .  We evaluate this uncertainty 
by varying the \met\ cutoff used to define the background sample 
(we use the \met\ ranges 0--5, 0--10, 0--15, and 10--15 \GeV ), 
and by using different parent and child requirements. 

We define our parent and child samples by varying the shower shape
requirements and by
tightening the selection by requiring the \dedx\ measured
in the tracking system to be consistent with that of an electron.  
Tables~\ref{tab:fzqcd1} and~\ref{tab:fzqcd2} show the results.

\begin{table}[ht!]
\begin{center}
\caption{The fraction of the \wev\ events in the CC that come from
multijet, $b$ quark, and direct photon sources, $f_{QCD}^{W\ {\rm CC}}$.  
In this table, ISO refers to the electron
isolation requirement, EMF to the requirement that the fraction of the
electron energy in the hadronic calorimeter be small, CHI refers to the
shower shape requirement, nominal means the electron identification criteria
used in the
\wev\ sample (CHI$<$100, ISO$<$0.15, and EMF$>$ 0.95, see Ch.3 of 
Ref.~\protect\cite{jamalsthesis}). \dedx\ means the matching track was 
required to have $dE/dx<1.4$ or 
$dE/dx>3.0$ for CDC tracks and $dE/dx<1.3$ or $dE/dx>2.5$ for FDC tracks, 
to reject photon conversions (see Ref.~\protect\cite{topprd}). } 
\medskip
\begin{tabular}{|c|c|c|c|c|} \hline
Parent cuts          &     Child Cuts &  $\epsilon_s$  & $\epsilon_b$ &    
$f_{QCD}^{W\ {\rm CC}}$ [\%]    \\
\hline
nominal    &   +\dedx\               &  0.933 $\pm$ 0.004 & 0.372 $\pm$ 0.023  
&     3.39$\pm$0.9  \\
ISO(0.15),EMF(0.95)   &     nominal  & 0.952 $\pm$ 0.003 & 0.686 $\pm$ 0.017
&            4.49$\pm$1.0  \\
EMF(0.95)            &     nominal  & 0.949 $\pm$ 0.003 & 0.650 $\pm$ 0.011 
&            4.41$\pm$0.8\\
EMF(0.9)             &     nominal  & 0.941 $\pm$ 0.004 & 0.573 $\pm$ 0.015 
&            4.46$\pm$0.8 \\
EMF(0.9),ISO(0.15)   &     nominal  & 0.945 $\pm$ 0.004 & 0.621 $\pm$ 0.016 
&            4.76$\pm$0.8 \\
EMF(0.9),CHI(100)    &     nominal  & 0.989 $\pm$ 0.003 & 0.872 $\pm$ 0.007 
&            6.16$\pm$2.0 \\
ISO(0.15),CHI(100)   &     nominal  & 0.992 $\pm$ 0.002 & 0.934 $\pm$ 0.007 
&            6.96$\pm$3.0 \\
\hline 
\end{tabular}
\label{tab:fzqcd1}
\end{center}
\end{table}

\begin{table}[ht!]
\begin{center}
\caption{
The fraction of the \wev\ events in the EC that come
from multijet, $b$ quark, and direct photon sources, $f_{QCD}^{W\ {\rm EC}}$. 
In this table, ISO refers to the electron isolation
requirement, EMF to the requirement that the fraction of the electron
energy in the hadronic calorimeter be small, CHI refers to the shower shape
requirement, nominal means the electron identification criteria used in the 
\wev\ sample (CHI$<$100, ISO$<$0.15, and EMF $>$ 0.95, see Ch. 3 of 
Ref.~\protect\cite{jamalsthesis}).  \dedx\ means
the matching track was required to have $dE/dx<1.4$ or $dE/dx>3.0$ for CDC 
tracks and $dE/dx<1.3$ or $dE/dx>2.5$ for FDC tracks 
(see Ref.~\protect\cite{topprd}). } \medskip
\begin{tabular}{|c|c|c|c|c|} \hline
Parent cuts          &     Child Cuts & $\epsilon_s$ & $\epsilon_b$ &    
$f_{QCD}^{W\ {\rm EC}}$ [\%] \\
\hline
nominal    &               +\dedx\   & 0.759 $\pm$ 0.010 & 0.552 $\pm$ 0.007 
&                  14.60$\pm$4.5\\
ISO(0.15),EMF(0.95)   &     nominal  & 0.881 $\pm$ 0.009 & 0.513 $\pm$ 0.016 
&            11.03$\pm$1.5\\
EMF(0.95)            &     nominal  & 0.880 $\pm$ 0.009 & 0.492 $\pm$ 0.015 
&            11.44$\pm$1.2\\
EMF(0.9)             &     nominal  & 0.868 $\pm$ 0.010 & 0.367 $\pm$ 0.015
&           14.48$\pm$1.2\\
EMF(0.9),ISO(0.15)   &     nominal  & 0.868 $\pm$ 0.009 & 0.398 $\pm$ 0.017 
&           14.13$\pm$1.4\\
EMF(0.9),CHI(100)    &     nominal  & 0.987 $\pm$ 0.004 & 0.858 $\pm$ 0.013 
&           19.99$\pm$3.3 \\
ISO(0.15),CHI(100)   &     nominal  & 0.991 $\pm$ 0.003 & 0.897 $\pm$ 0.010 
&           21.94$\pm$3.8 \\
\hline 
\end{tabular}
\label{tab:fzqcd2}
\end{center}
\end{table}

The uncertainty on the background fraction is dominated by the
uncertainties on $\epsilon_s$ and $\epsilon_b$, and is given
approximately by
\begin{equation}
\delta f^W_{QCD} \approx \frac{\epsilon_s}{\epsilon_s-\epsilon_b} \delta
\epsilon_s
\oplus
\frac{\epsilon_b f^W_{QCD}}{\epsilon_s-\epsilon_b}
\delta \epsilon_b.
\end{equation}
From this equation, one can see that the method works best when
$\epsilon_s - \epsilon_b$ is large, and produces large errors when
this difference is small. 
We take a Gaussian distribution (normalized to unity) with the mean and 
uncertainty corresponding to each background fraction in 
Tables~\ref{tab:fzqcd1} and~\ref{tab:fzqcd2}. For the mean value of 
$f^W_{QCD}$, we add all the CC or EC distributions and take the median of the 
resulting distribution. We set the systematic uncertainty in $f^W_{QCD}$ from 
the symmetric band around the median with an area of 68\% 
of the total distribution.
The results are $f^{W\ {\rm CC}}_{QCD} = \wfqcdcc \pm \wfqcdccerr$ for \CC\ 
events, 
and $f^{W\ {\rm EC}}_{QCD} = \wfqcdec \pm \wfqcdecerr$ for \EC\ events.  
To obtain the combined background fraction, we combine the \CC\ and \EC\ $W$
boson cross sections. The weights for \CC\ and \EC\ events are 
taken as $1/\delta_{u}^{2}$ , where $\delta_{u}$ is the total uncorrelated 
error for each individual cross section, and where we make the conservative 
assumption that there is maximal correlation between the \CC\ and \EC\ 
uncertainties (the correlated part for each uncertainty
is the smaller of the two). We then find the background fraction that 
corresponds to this combined $W$ boson cross section.
The combined background fraction is estimated to be 
$f^W_{QCD} = \wfqcd \pm \wfqcderr$.  

\begin{figure}[hp!]
\center
\centerline{\psfig{figure=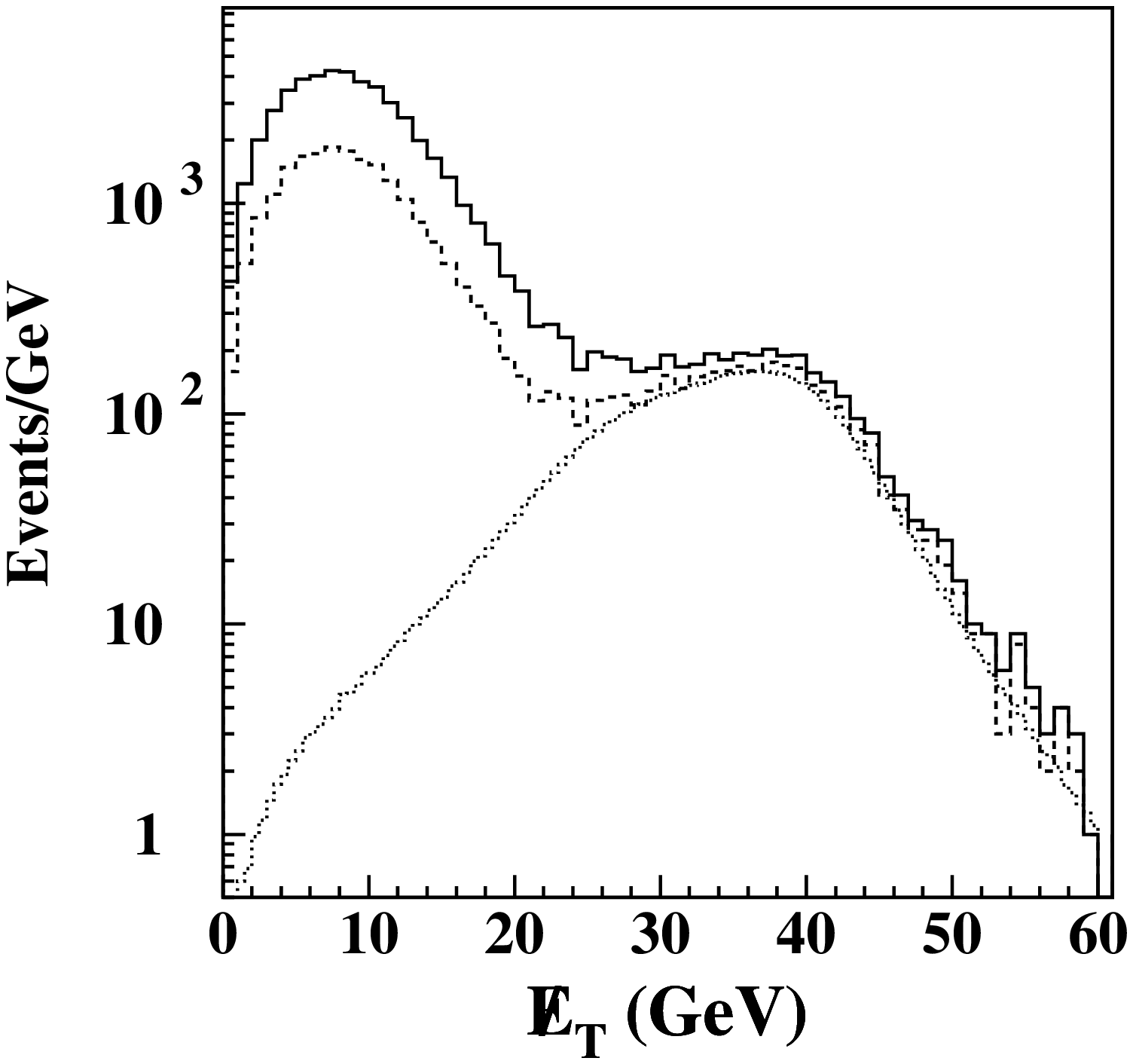,height=\figsize}}
        \caption{
The \met\ distribution for a particular choice of parent and child inclusive
electron samples. 
The solid line is the parent sample corresponding to the nominal \wb\ boson 
selection cuts except for \met. The dashed line is the child sample, 
corresponding to nominal cuts and the additional \dedx\ requirement. The 
dot-dash line is the sum of \wev\ and \wte\ from Monte Carlo. There is 
negligible $W$ boson contribution in the low \met\ background regions.
}  \label{fig:met_dist}
\end{figure}

The method we use to obtain $\epsilon_b$ assumes that the efficiency
for background events to pass the electron identification requirements
is the same for events with small and large \met. Most of our identification
requirements are calorimeter-based and can, in principle,  be
correlated with \met. However, the $f^W_{QCD}$ measurement obtained
by adding the tracking-based \dedx\ requirement yields results consistent
with the calorimeter-based methods, giving us confidence that
the correlations between \met\ and $\epsilon_b$ are small.
Our studies assume the contamination from the $W$ boson in the backgrounds is 
small in the low \met\ region. 
We check the validity of this assumption by looking at
the \met\ distributions of the child and parent samples and compare them to 
the \met\ distribution from \wev\ and \wte\ Monte Carlo events. 
Figure~\ref{fig:met_dist} shows the case where the parent background sample 
corresponds to the nominal $W$ boson selection (except
for the \met\ requirement) and the child sample is obtained from the 
additional \dedx\ requirement.  
The Monte Carlo 
distribution is normalized to the background sample distributions in the high
\met\ region, which is dominated by real $W$ boson events. 
The fraction of $W$ boson events in the low \met\ region is found to be 
negligible. 

\subsection{Backgrounds from multijet, \mbox{\boldmath $b$} quark, and
direct photon sources in the \mbox{\boldmath \zee\ } sample}

\begin{figure}[hp!]
\center
\centerline{\psfig{figure=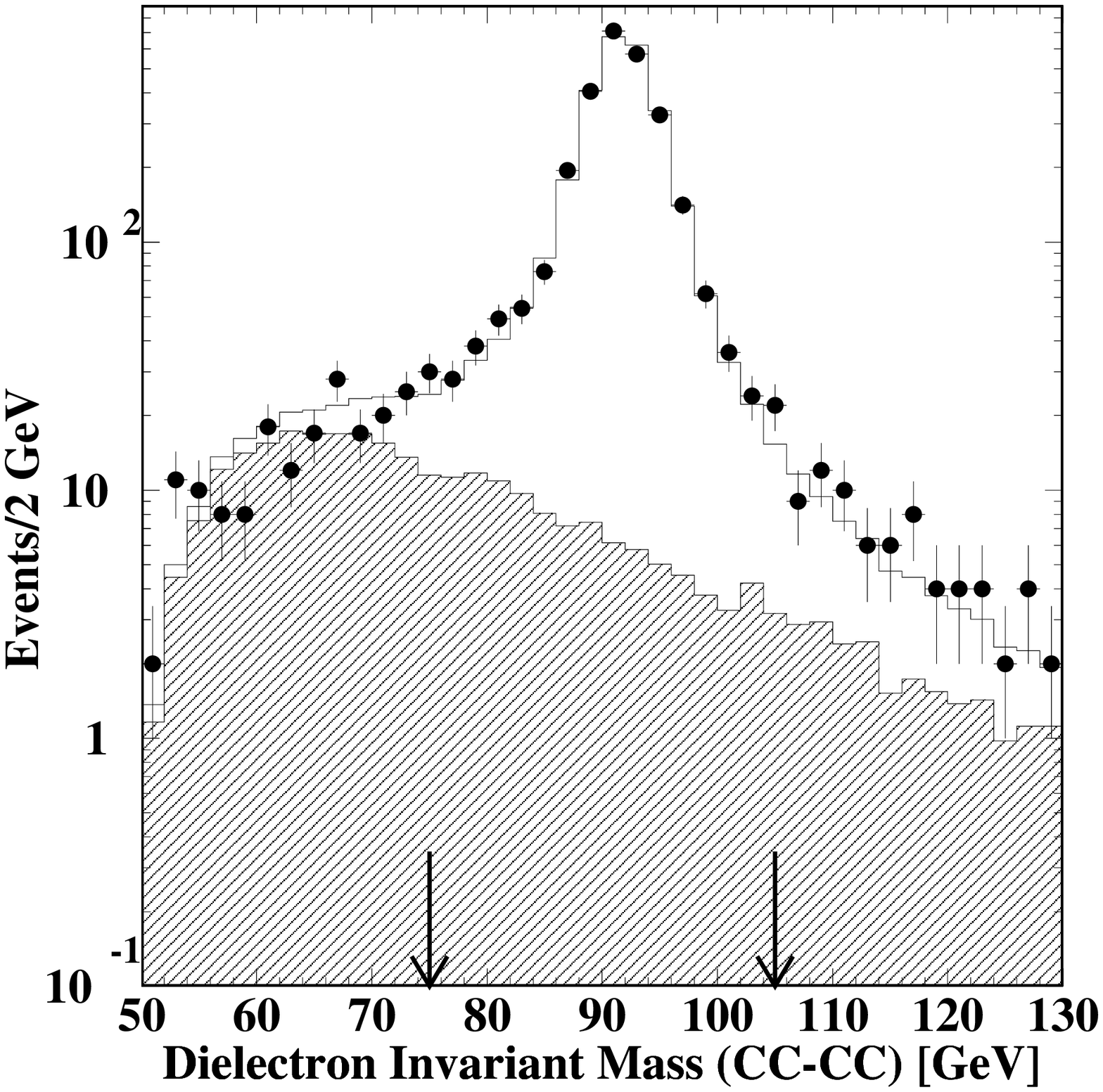,height=\figsize}}
        \caption{
Fit of the \zee\ invariant mass distribution. The shaded histogram 
is the background shape obtained from direct photon data, and the dots are the 
\zee\ candidates. The solid line histogram results from fitting the data to a  
linear combination of the Drell-Yan signal shape from \PYTH\ 
and the background shape.  
}  \label{fig:zfitcccc}
\end{figure}

\begin{figure}[hp!]
\center
\centerline{\psfig{figure=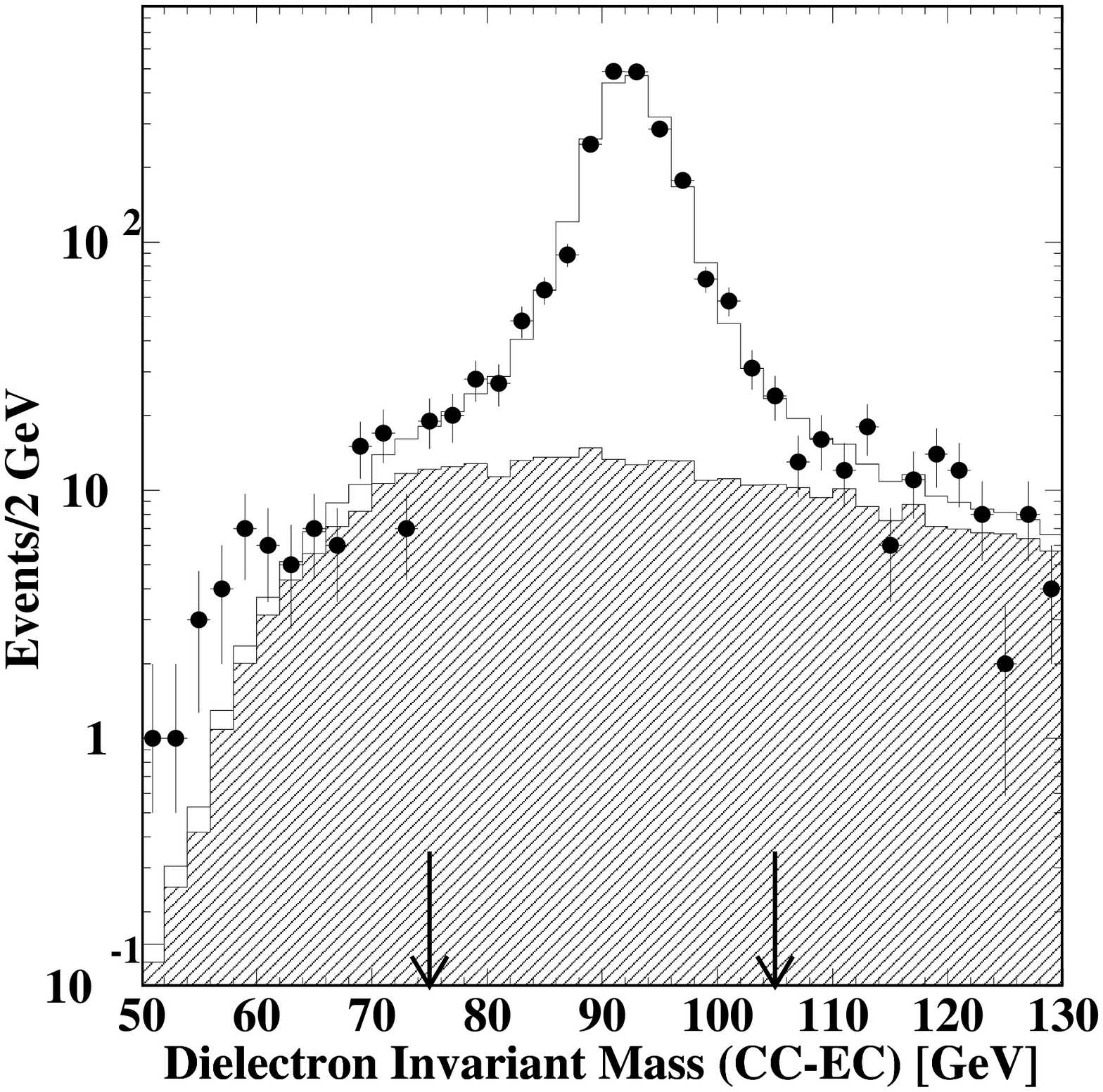,height=\figsize}}
        \caption{
Fit of the \zee\ invariant mass distribution. The shaded histogram 
is the background shape obtained from direct photon data, and the dots are the 
\zee\ candidates. The solid line histogram results from fitting the data to a  
linear combination of the Drell-Yan signal shape from \PYTH\ 
and the background shape.  
}  \label{fig:zfitccec}
\end{figure}

\begin{figure}[hp!]
\center
\centerline{\psfig{figure=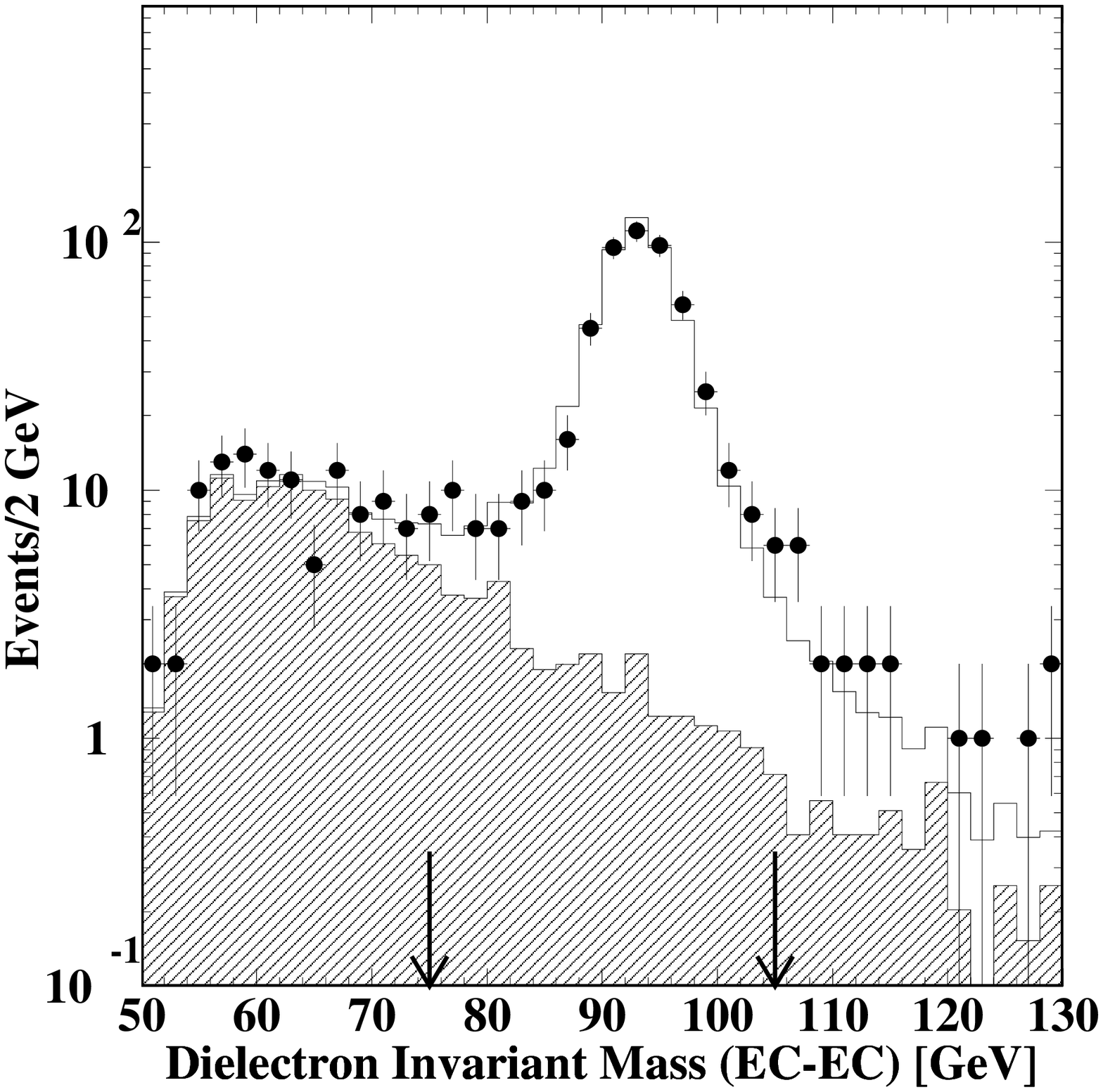,height=\figsize}}
        \caption{
Fit of the \zee\ invariant mass distribution. The shaded histogram 
is the background shape obtained from direct photon data, and the dots are the 
\zee\ candidates. The solid line histogram results from fitting the data to a  
linear combination of the Drell-Yan signal shape from \PYTH\ 
and the background shape.  
}  \label{fig:zfitecec}
\end{figure}

The background fraction for the \zee\ sample due to multijet, $b$ quark, 
and direct photon sources is determined by fitting the dielectron invariant 
mass distribution to a linear combination of a signal shape, obtained from
$Z/\gamma^{*}$ events generated with \PYTH\ and processed 
through the detector simulation, and a background shape determined
from data. Different mass distributions from different sources, such as 
multijet events, direct photon candidates, and events passing all of the 
\zee\ kinematic cutoffs, but failing the electron identification requirements, 
are used for background shapes. 
Figures~\ref{fig:zfitcccc},~\ref{fig:zfitccec}, and~\ref{fig:zfitecec} show 
such fits with 
a background shape determined from direct photon data, for the case
where both electrons are in the \CC, for the case where one electron is
in the \CC\ and the other in the \EC, and for the case where both electrons
are in the \EC, respectively. 
Systematic uncertainties are determined from the range of values obtained using
the different background shapes and by varying the range of
invariant masses used in the fit.  The result is $f^Z_{QCD} = \zfqcd \pm
\zfqcderr$.

\subsection{\mbox{\boldmath $W$ and $Z$ Boson Backgrounds in the \wev\ Sample}}

The other sources of background in the \wev\ sample are 
\zee, \ztt, and \wtv\ events.
A \zee\ event can be misidentified as a \wev\ event when one of the
electrons fails the fiducial requirements or is misidentified as a
jet, and the transverse energy in the event is substantially
mismeasured, yielding a  large apparent \met.  
Events from the process \ztt\ can also mimic \wev\ events.
\wtv\ events, in which the $\tau$ decays to an electron, 
are identical to \wev\ events, except that on average the electron \et\ and 
the \met\ are lower.
The size of these backgrounds scales with the \wev\ or \zee\ cross 
section, and this must be taken into account when the
background subtraction is done. 
To take this into account, we determine \sigmaw\ from the relationship
\begin{eqnarray}
N_{W} &=& N^{W}_{obs}(1-f^{W}_{QCD}) = N_{e}+N_{\tau} + N_{Z}^{W} \nonumber \\
      &=&\left[ \epsilon_W \cdot A_W \cdot {\sigma(p \overline{p} 
\rightarrow W+X) \cdot B(W \rightarrow e \nu)} \cdot {\cal L } \right]
 + 
\label{eqn:bige}  \\
& &\left[ \epsilon_W \cdot A_{W\tau}^W \cdot {\sigma(p \overline{p} 
\rightarrow W+X)
\cdot B(W \rightarrow \tau \nu)} \cdot {\cal L } \right] + N_{Z}^{W} 
\nonumber
\end{eqnarray}
where $N_W$ is the number of candidate \wb\ boson events after correcting for
backgrounds from  multijets, direct photons, and $b$ quarks;
$N^{W}_{obs}$ is the number of candidate \wev\ events;
$N_{e}$ is the number of \wev\ events passing the \wev\ selection criteria;
$N_{\tau}$ and $N_{Z}^{W}$ are the numbers
of \wtv\ and \zee\ events respectively passing these criteria;
$A_{W\tau}^W$ is the fraction of the \wtv\ events that
passes the \wev\ selection criteria; and
$\cal L$ is the integrated luminosity. 
We assume in Eq.~\ref{eqn:bige} that the \wb\ boson couples with equal 
strength to all lepton flavors, and therefore
$B(W \rightarrow \tau \nu) = B(W \rightarrow e \nu)$.  

The  \zee\ and \ztt\ backgrounds are estimated using a \GEAN-based simulation 
of the detector, with \HERW~\cite{herwig} to generate both
\zee\ and \ztt\ events. 
The number of \zb\ boson background events in the \wev\ sample is estimated by 
\begin{equation}
N^W_Z = 
\epsilon_W \cdot N_{obs}^Z (1-f_{QCD}^Z) \cdot 
\frac{A_{Zee}^W + A_{Z\tau}^W}
{A_Z \cdot \epsilon_Z} 
\end{equation}
where $A^W_{Zee}$ is the fraction of \zee\ events that passes the \wev\ 
selection criteria; 
$A^W_{Z\tau}$ is the fraction of \ztt\ events that passes the \wev\ 
selection criteria;
$N_{obs}^Z$ is the number of candidate \zee\ events; 
$f_{QCD}^Z$ is the fraction of these candidates from multijet, $b$ quark, and 
direct photon background sources;
$\epsilon_Z$ is the electron identification efficiency for \zee\ events; and 
$A_Z$ is the geometric and kinematic acceptance for \zee\ events.
The ratio $(A_{Zee}^W + A_{Z\tau}^W)/A_Z$
is found to be $\wacczinw \pm \wacczinwerr$, and thus a total of $621 \pm 155$ 
\zb\ boson events is expected to pass the \wev\ selection.
The uncertainty in this estimate has two main components: the difference 
between the electron identification efficiency in the simulation and in the 
data, and the effect of any additional overlapping minimum-bias events. This 
uncertainty has a negligible effect on the overall uncertainty in the \wb\ 
boson cross section and the ratio $\cal R$.

The backgrounds to the \wev\ and \zee\ samples from the decays
\wtv\ and \ztt, where $\tau \rightarrow e \nu$,
are calculated using the same \wb\ and \zb\ boson production and
decay model as in the acceptance calculation.  The tau leptons 
are forced to decay electronically and then the event is smeared.  
Backgrounds from $\tau$ in the \zee\ sample are
found to be negligible. Assuming lepton universality and  the
fact that we do not observe any dependence of the lepton identification
efficiency on the transverse energy of the lepton,
we can account for the $\tau$ backgrounds in the \wb\ 
boson sample by making a correction to the \wb\ boson acceptance of 
$(1 + \frac{A_{W\tau}^W}{A_W}) = \wtaucorr \pm \wacctauerr$.


\section{Luminosity}
\label{sec:luminosity}
A precise value of the integrated luminosity is needed for determining
any absolute cross section. 
This analysis uses data collected at \sqrts\ = 1.8 \TeV\ during the 
1994--1995 running of the Fermilab Tevatron. The measurement of luminosity is 
described in detail in Refs.~\cite{jamalsthesis,ggthesis}. 
The luminosity ($L$) is related to the counting rate in the 
\lzero\ counters ($R_{\lzero}$) by~\cite{lum_comment}
\begin{equation}
L = \frac{-\ln(1-\tau R_{\lzero})}{\tau\sigma_{\lzero}}
\end{equation}
where $\sigma_{\lzero}$ is the effective \ppbar\ cross section subtended by 
the \lzero\ counters,
and $\tau = 3.5\ \mu s$ is the time interval between beam crossings.
$R_{\lzero}$ is defined by the counts observed in six trigger scalers, one for
each beam bunch, divided by the fixed time between crossings. This counting 
rate never saturated during the run, not even at the highest  
luminosities. 
Assuming Poisson statistics, a correction is applied to account for 
multiple interactions. The value of $\sigma_{\lzero}$ is obtained from
\begin{equation}
\sigma_{\lzero} = \epsilon_{\lzero}^{\ppbar}(A_{sd}\sigma_{sd} +
A_{dd}\sigma_{dd} + A_{nd}\sigma_{nd})
\end{equation}
where the single diffractive ($\sigma_{sd}$), double diffractive 
($\sigma_{dd}$), and non diffractive ($\sigma_{nd}$) components of the total 
inelastic \ppbar\ cross section are combined into a ``world average'' using 
the results from CDF~\cite{lum_cdf}, E710~\cite{lum_e710}, 
and E811~\cite{lum_e811}; 
the \lzero\ trigger efficiency $\epsilon_{\lzero}^{\ppbar}$ is determined 
using samples of data collected from triggers on random beam crossings; and 
the different \lzero\ acceptances 
($A_{sd}$, $A_{dd}$, $A_{nd}$) are obtained from Monte 
Carlo studies. Table~\ref{tab:lumtable} shows the inputs to our 
calculation of $\sigma_{\lzero}$. 

\begin{table}[ht!]
\begin{center}
\caption{Values used in the $\sigma_{\lzero}$ calculation; SD, DD and ND 
refer to single diffractive, double diffractive, and non diffractive, 
respectively.}
\medskip
\begin{tabular}{|l|c|}
\hline 
SD Acceptance ($A_{sd}$)       & 15.1\% $\pm$ 5.5\% \\
DD Acceptance ($A_{dd}$)       & 71.6\% $\pm$ 3.3\% \\
ND Acceptance ($A_{nd}$)       & 97.1\% $\pm$ 2.0\% \\
\hline
\lzero\ Trigger Efficiency ($\epsilon_{\lzero}^{\ppbar}$) & 91\% $\pm$ 2\% \\
\hline
SD Cross Section ($\sigma_{sd}$)      &  9.54 mb $\pm$ 0.43 mb \\
DD Cross Section ($\sigma_{dd}$)      &  1.29 mb $\pm$ 0.20 mb \\
ND Cross Section ($\sigma_{nd}$)      & 46.56 mb $\pm$ 1.63 mb \\
\hline
$\sigma_{\lzero}$                     & 43.1  mb $\pm$ 1.9  mb \\
\hline 
\end{tabular}
\label{tab:lumtable}
\end{center}
\end{table}

\begin{figure}[ht!]
\center
\centerline{\psfig{figure=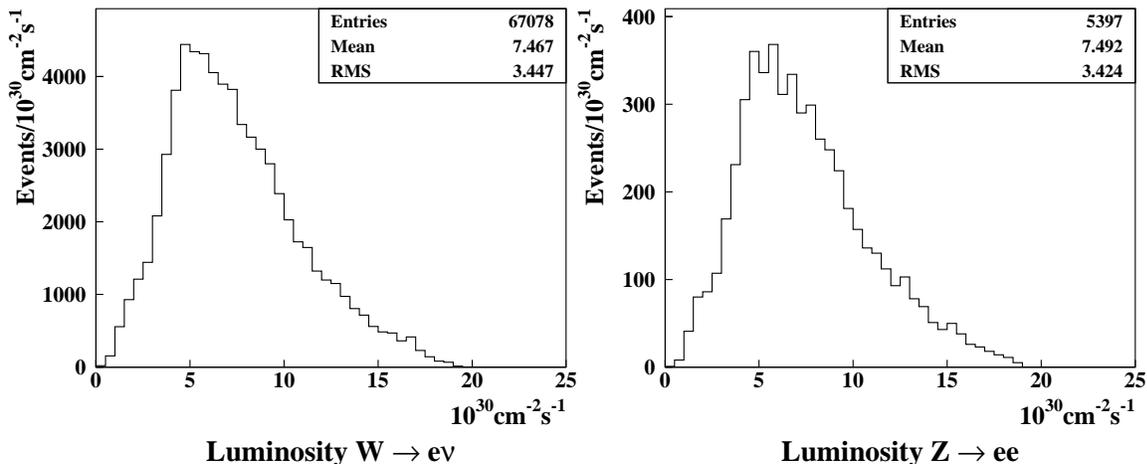,width=\linewidth}}
        \caption{
Distribution in luminosity for \wev\ or \zee\ candidates. The 
mean and RMS values of the distributions are consistent with each other.
}  \label{fig:instlum}
\end{figure}

Luminosities during the 1994--1995 running
period ranged from 2--20 $\times 10^{30}$ cm$^{-2}$s$^{-1}$.
The average luminosity for the \wev\ and \zee\ data  
samples is $7.5 \times 10^{30}$ cm$^{-2}\rm{s}^{-1}$, with an average 
of 1.6 interactions per beam crossing. The integrated luminosity for the 
\zee\ and \wev\ data samples is \lumb\ $\pm$ \lumberr\ \ipb. The uncertainty 
in luminosity is the dominant uncertainty in the measurement of \wb\ and \zb\ 
boson cross sections.
Figure~\ref{fig:instlum} shows the distribution in luminosity 
at the time of recording of the \wev\ and \zee\ candidates. 

It should be noted that CDF and previous \Dzero\ measurements used different
normalizations for luminosity.
The CDF Collaboration bases its luminosity purely on its own 
measurement of the inelastic \ppbar\ cross section~\cite{lum_cdf,lum_cdf_tm}. 
As a result, current luminosities used by CDF are 6.2\% lower than those 
used by \Dzero, and consequently all \Dzero\ cross sections are 
{\it ab initio} 6.2\% lower than all CDF cross sections.
Previous \Dzero\ measurements relied only on results from CDF and E710. 
Including the recent E811 measurement of the inelastic \ppbar\ cross section 
in the world average increased the discrepancy in normalization relative to
CDF from 3.0\% to 6.2\% (i.e., current values are 3.2\% higher than previous
\Dzero\ measurements). 
The luminosity measurement used by \Dzero\ prior to the E811 result is 
described more extensively in Ref.~\cite{lum_tech}.


\section{The Cross Sections and their Ratio}
\label{sec:crosssectionresults}
The product of the $W$ boson cross section and the
branching fraction for \wev\ is calculated using the relation
\begin{equation}
\sigma(p \overline{p} \rightarrow W + X) \cdot B(W \rightarrow e \nu) =
\frac{N_{obs}^W \cdot \left( 1-f_{QCD}^W \right) - 
\epsilon_W \cdot N_{obs}^Z (1-f_{QCD}^Z) \cdot 
\frac{A_{Zee}^W + A_{Z\tau}^W}
{A_Z \cdot \epsilon_Z} }
{
\epsilon_W \cdot A_{W} \cdot 
\left( 1+\frac{A_{W\tau}^W}{A_W} \right)
\cdot {\cal L}}  
\label{eqn:wcross}
\end{equation}
where $N_{obs}^W$ and $N_{obs}^Z$ are the number of \wev\ and \zee\ candidate 
events, respectively; 
$f_{QCD}^W$ and $f_{QCD}^Z$ are the fraction of the \wev\ and \zee\ candidate 
events, respectively, that come from
multijet, $b$ quark, and direct photon background sources; 
$\epsilon_W$ and $\epsilon_Z$ are the efficiency for \wev\ and \zee\ events,
respectively, to pass the selection requirements; 
$A_W$ and $A_Z$ are the geometric and kinematic acceptance for \wev\ and 
\zee, respectively, which include effects from detector resolution; 
$A_{W\tau}^W$, $A_{Zee}^W$ and $A_{Z\tau}^W$ are the fraction of \wtv, \zee, 
and \ztt\ events, respectively, that passes the \wev\ selection 
criteria;
and $\cal L$ is the integrated luminosity of the data sample.  

The product of the $Z$ boson cross section and the branching fraction for 
\zee\ is determined from the relation
\begin{equation}
\sigma(p\overline{p} \rightarrow Z+X)\cdot B(Z \rightarrow ee) =
\frac{N_{obs}^Z \cdot \left( 1 - f_{QCD}^Z \right) \cdot 
\left( 1 -f_{DY}\right)
}{\epsilon_Z \cdot A_Z \cdot {\cal L}}
\label{eqn:zcross}
\end{equation}
where $f_{DY}$ is a correction for the Drell-Yan contribution to
$Z$ boson production. The ratio ${\cal R}$ can therefore be written as
\begin{eqnarray}
\nonumber
{\cal R} &=&
\frac{\epsilon_Z}{\epsilon_W} \cdot \frac{A_Z}{A_W} \cdot
\frac{1}{1+\frac{A^W_{W\tau}}{A_W}} \cdot
\frac{1}{1-f_{QCD}^Z}  \cdot
\frac{1}{1-f_{DY}}  \\
  & \times & 
\left[
\frac{N_{obs}^W}{N_{obs}^Z} \cdot \left( 1-f_{QCD}^W \right) -
\epsilon_W \cdot
\frac{(A_{Zee}^W+A_{Z\tau}^W)  \cdot \left( 1-f_{QCD}^Z \right)}
{A_Z \cdot \epsilon_Z}
\right]
\label{eqn:ratio}
\end{eqnarray}

The uncertainties on the individual cross sections are dominated by
the uncertainty on the integrated luminosity measurement (4.3\%).  
Tables~\ref{tab:wcross} and~\ref{tab:zcross} 
summarize the results for the individual cross sections.
The result for \sigmaw\  is \wxsec\ $\pm$ \wxstat\ (stat)
                              $\pm$ \wxsyst\ (syst)
                              $\pm$ \wxlum\ (lum) \pb.
The result for \sigmaz\ is  \zxsec\ $\pm$ \zxstat\ (stat)
                             $\pm$ \zxsyst\ (syst)
                             $\pm$ \zxlum\ (lum)  \pb.

\begin{table}[!h]
\begin{center}
\caption{Values used in the \wev\ cross section measurement. }
\medskip
\begin{tabular}{ccc}
\sigmaw\   & \wxsec\ $\pm$ \werr\ \pb & \\
           &                          & \\
\hline
           &                          & \\
           & Value & Uncertainty Contribution [\pb] \\ 
\hline
${N_{obs}^W}$  & \wnum                 & 10   \\
${\epsilon_W}$ & \weff\ $\pm$ \wefferr & 30  \\
${A_W}$ & \wacc\ $\pm$ \waccerr        & 20  \\
$f_{QCD}^W$    & \wfqcd\ $\pm$ \wfqcderr  & 35  \\
$(A_{Zee}^W + A_{Z\tau}^W)/A_Z$ & \wacczinw\ $\pm$ \wacczinwerr & -- \\
${\epsilon_Z}$ & \zeff\ $\pm$ \zefferr    & --  \\
$f_{QCD}^Z$    & \zfqcd\ $\pm$ \zfqcderr  & --   \\
$N_Z^W$        & $621 \pm 155$            & 6    \\
$A_{W\tau}^W/A_W$ & 0.0211 $\pm$ 0.0021   & 5   \\ 
$\cal L$          & \lumb\ $\pm$ \lumberr\ \ipb & 100 \\
\end{tabular}
\label{tab:wcross}
\end{center}
\end{table}

\begin{table}[!h]
\begin{center}
\caption{Values used in the \zee\ cross section measurement. }
\medskip
\begin{tabular}{ccc}
\sigmaz\      & \zxsec\ $\pm$ \zerr\ \pb & \\
              &                     & \\
\hline
           &                          & \\
           & Value & Uncertainty Contribution [\pb] \\ 
\hline
${N_{obs}^Z}$  & \znum  &  3  \\
${\epsilon_Z}$ & \zeff\ $\pm$ \zefferr & 3  \\
${A_Z}$        & \zacc\ $\pm$ \zaccerr & 2  \\
$f_{QCD}^Z$    & \zfqcd\ $\pm$ \zfqcderr & 1    \\
$f_{DY}$       & \zfdy\ $\pm$ \zfdyerr & $< 1$  \\
$\cal L$       & \lumb\ $\pm$ \lumberr\ \ipb & 10 \\
\end{tabular}
\label{tab:zcross}
\end{center}
\end{table}

\begin{figure}[hp!]
\center
\centerline{\psfig{figure=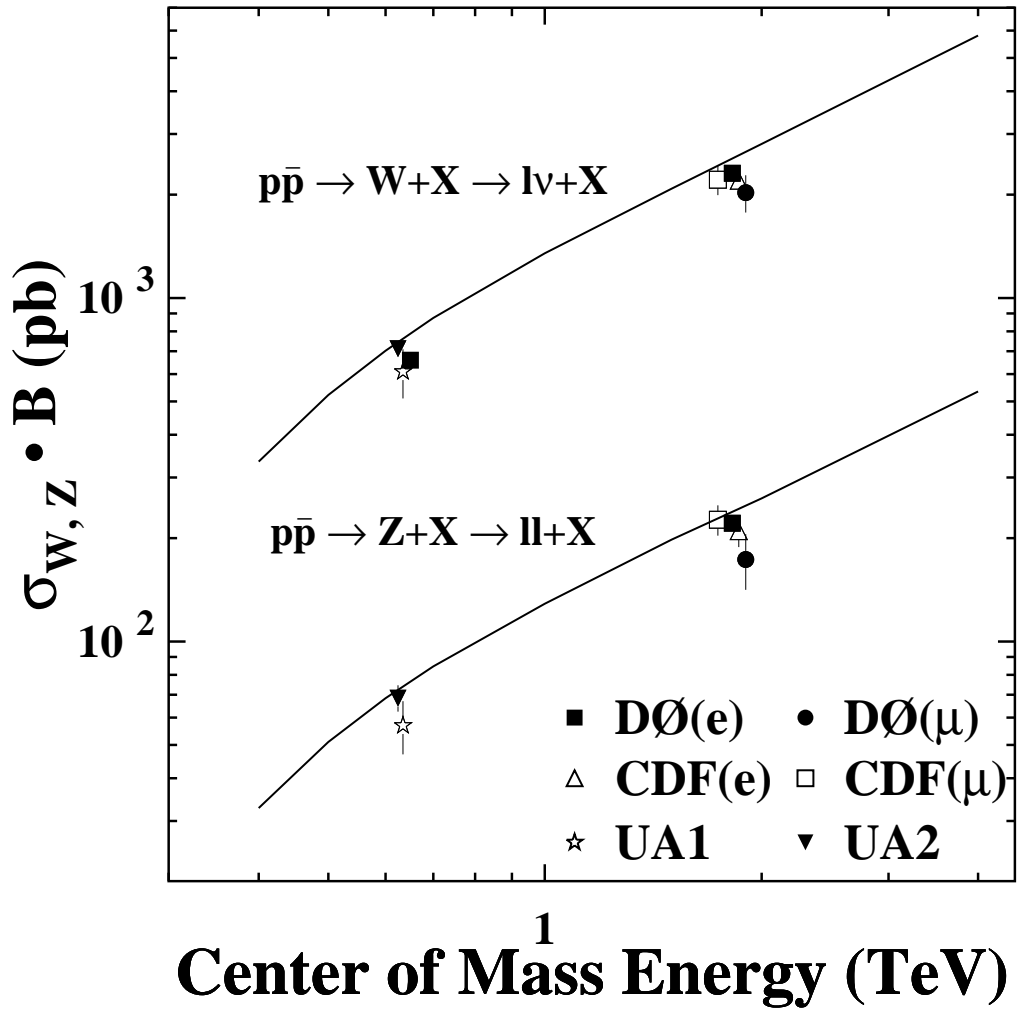,height=150mm}}
        \caption{
Comparison between measured and predicted cross sections. 
The lines correspond to a theoretical calculation of order $\alpha_s^2$
using the program of Ref.~\protect\cite{theoryrs} 
with the CTEQ4M structure functions, a \zb\ boson mass of 91.188 \GeV,
a \wb\ boson mass of 80.375 \GeV, and $\sin^{2}{\theta_W}$=0.2231.
The \Dzero\ results in the muon channel are
from Ref.~\protect\cite{d01aR} normalized to the new luminosity. 
}  \label{fig:crosssectionpic}
\end{figure}

\begin{figure}[hp!]
\center
\centerline{\psfig{figure=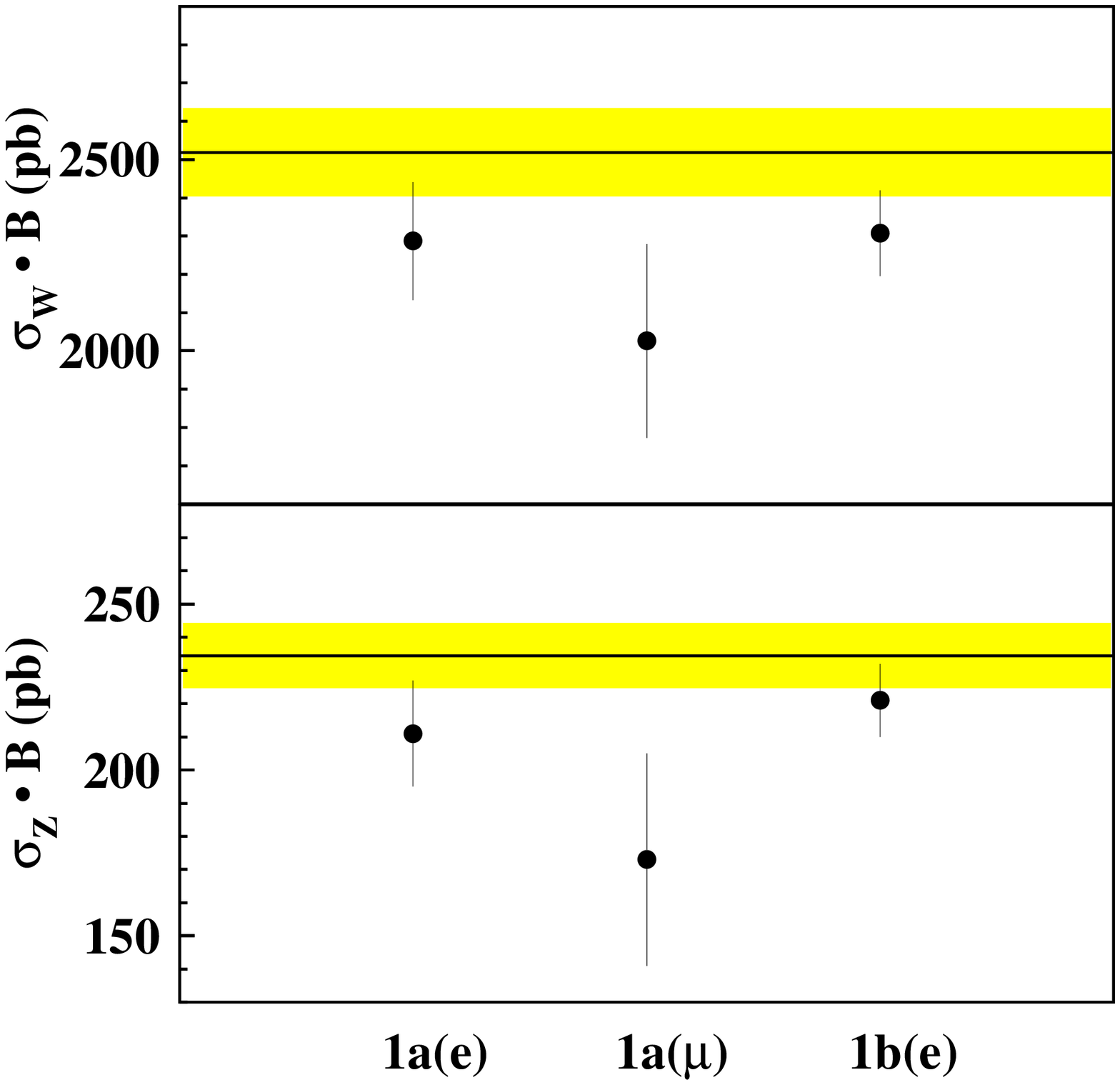,height=\figsize}}
        \caption{
Run 1a (1992--1993)~\protect\cite{d01aR} and 1b (1994--1995) results for the 
\wb\ and \zb\ boson cross sections times branching fractions. 
The line is the theoretical prediction from Ref.~\protect\cite{theoryrs}. The 
central value uses $\Lambda_{\rm QCD} = 296$ MeV and the CTEQ4M structure 
functions. The shaded region shows the uncertainty in the prediction due to 
variations in $\alpha_s$ obtained by varying $\Lambda_{\rm QCD}$ between 
213 MeV and 399 MeV. The Run 1a results have 
been normalized to the new luminosity to be consistent with Run 1b results.
}  \label{fig:cross2}
\end{figure}

Figure~\ref{fig:crosssectionpic} shows a comparison between our results and  
calculations of order $\alpha_s^2$ using the program of Ref.~\cite{theoryrs} 
with the CTEQ4M structure functions, a \zb\ boson mass of 91.188 \GeV,
a \wb\ boson mass of 80.375 \GeV, and $\sin^{2}{\theta_W}$=0.2231. The 
\Dzero\ results in the muon channel~\cite{d01aR} are from Run 1a(1992--1993), 
and have been multiplied by 0.969 for consistency with the new luminosity 
normalization. 
Figure~\ref{fig:cross2} shows the Run 1b (1994--1995) results for the 
individual \wb\ and \zb\ boson cross sections times electronic branching 
fraction and the previous \Dzero\ results from Run 1a 
(1992--1993)~\cite{d01aR} for both the electron and muon channels compared to 
the corresponding theoretical predictions.
The Run 1a results are normalized to the new luminosity for consistency with 
Run 1b results.

Table~\ref{tab:theratio} summarizes the result for
the ratio of the cross sections, \Reqn. 
\begin{table}[h!]
\centering
\caption{Values used in the Ratio Measurement.}
\begin{tabular}{ccc}
 $\cal R$      & \rxsec\ $\pm$ \rerr & \\
               &                     & \\
\hline
    &         & \\
    &  Value  &   Uncertainty Contribution \\
\hline
 $N_{obs}^W/N_{obs}^Z$ & \rnum\ $\pm$ \rnumerr &  0.15\\
 $\epsilon_Z/\epsilon_W$ & \reff\ $\pm$ \refferr & 0.06 \\
 $A_{Z}/A_W$ & \racc\ $\pm$ \raccerr &  0.09\\
 $(A_{Zee}^{W}+A_{Z\tau}^{W})/A_Z$ & \wacczinw\ $\pm$ \wacczinwerr & 0.03\\
 $f_{QCD}^W$ & \wfqcd\ $\pm$ \wfqcderr & 0.16\\
 $f_{QCD}^Z$ & \zfqcd\ $\pm$ \zfqcderr & 0.05 \\
 $f_{DY}$ & \zfdy\ $\pm$ \zfdyerr & 0.01\\
 $A_{W\tau}^W/A_W$ & \wacctau\ $\pm$ \wacctauerr & 0.02\\ 
\end{tabular}
\label{tab:theratio}
\end{table}
In the ratio, many of the systematic uncertainties, including the
luminosity uncertainty, cancel. The uncertainty in ${\cal R}$ has
five main components: 
the uncertainty in the multijet, $b$ quark, and direct photon backgrounds
to the \wb\ boson (\rerrwqcd); 
the statistics of the $Z$ boson sample (\rerrzstat);
the uncertainty in the ratio of the $W$ and $Z$ boson acceptances (\rerracc); 
the uncertainty in the ratio of the $W$ and $Z$
boson electron identification efficiencies (\rerreff); 
and the uncertainty in the multijet, $b$ quark, and
direct photon backgrounds to the \zb\ (\rerrzqcd).
In addition, we assign a 1\% uncertainty in $\cal R$ due to 
next-to-leading-order electroweak radiative corrections.
The result is $\cal{R} = $ \rxsec\ $\pm$ \rxstat\ (stat)
                $\pm$ \rxsyst\ (syst) $\pm$ \rxnlo\ (NLO).



\section{Consistency Checks}
\label{sec:checks}
\subsection{Cross Sections from the Individual Cryostats}

As a consistency check, we calculate the \wb\ and \zb\ boson cross sections
using the data from each calorimeter cryostat individually and compare the 
differences between them with the uncorrelated uncertainties. The luminosity
uncertainty is 100\% correlated between the different cyostats and therefore
is not used in these comparisons.
For the \CC\ alone, the result for \sigmaw\  is
2308 $\pm$ 11 (stat) $\pm$ 51 (syst) $\pm$ 99 (lum) \pb . For the \EC , 
the result is 2207 $\pm$ 16 $\pm$ 121 $\pm$ 95 \pb. 
The dominant uncertainties in the \CC\ are 
the uncertainty on the acceptance ($\pm$ 21 \pb);
the uncertainty on the efficiency  ($\pm$ 31 \pb);
and the uncertainty from the multijet, $b$ quark, and direct photon
background ($\pm$ 34 \pb).
The dominant uncertainties in the \EC\ are 
on the acceptance ($\pm$ 20 \pb);
the efficiency  ($\pm$ 41 \pb);
and the  multijet, $b$ quark, and direct photon 
background  ($\pm$ 112 \pb).
The uncertainties in the acceptances come from uncertainties in the calorimeter
energy scales (mostly uncorrelated), assumptions on the distribution of 
\wb\ and \zb\ boson transverse momentum (correlated), and assumptions on the 
effects of final state radiation (correlated).  
The systematic uncertainties in the efficiencies are mostly 
correlated. There is a statistical component
that would be uncorrelated, but we neglect it here and assume
the efficiencies are correlated. The uncertainties in 
QCD backgrounds are mostly uncorrelated between the \CC\ and the \EC.
Using the full uncertainty in
the background (the uncertainties in acceptance 
and efficiency can be neglected for the purposes of this comparison),
we estimate the difference between the \CC\ 
and \EC\ cross sections as 
101 $\pm$ 19 $\pm$ 117 \pb.

Using only \CC-\CC\ combinations, the result for \sigmaz\ is
223 $\pm$ 4 $\pm$ 4 $\pm$ 10 \pb .
For \CC-\EC\ combinations, it is
216 $\pm$ 5 $\pm$ 4 $\pm$ 9 \pb .
For \EC-\EC\ combinations, it is
235 $\pm$ 10 $\pm$ 5 $\pm$ 10 \pb .
The dominant uncertainty in the \CC-\CC\  measurement is
from the uncertainty on the lepton identification efficiency (3.5 \pb).
The dominant uncertainties in the \CC-\EC\ measurement are from 
lepton identification (3.3 \pb) and QCD background (2.3 \pb). 
In the \EC-\EC\ measurements, lepton identification
contributes 4.5 \pb\ to the uncertainty, and QCD background contributes 
2.5 \pb.
To estimate the errors on the difference, we assume that 
the efficiencies are correlated.  For the \CC-\CC\ measurement, the background
contribution is small.  Because the \CC-\EC\ and \EC-\EC\ backgrounds 
both contain an \EC\ electron candidate, we assume the
background is 100\% correlated.  We therefore consider only the statistical
uncertainty, and we get
$\sigma_{{\rm CC-CC}}-\sigma_{{\rm CC-EC}} = 7 \pm 6$ \pb,
$\sigma_{{\rm CC-CC}}-\sigma_{{\rm EC-EC}} = -12 \pm 11$ \pb, 
and $\sigma_{{\rm CC-EC}}-\sigma_{{\rm EC-EC}} = -19 \pm 11$ \pb.

\subsection{Dependence on Instantaneous Luminosity}

\begin{figure}[hp!]
\center
\centerline{\psfig{figure=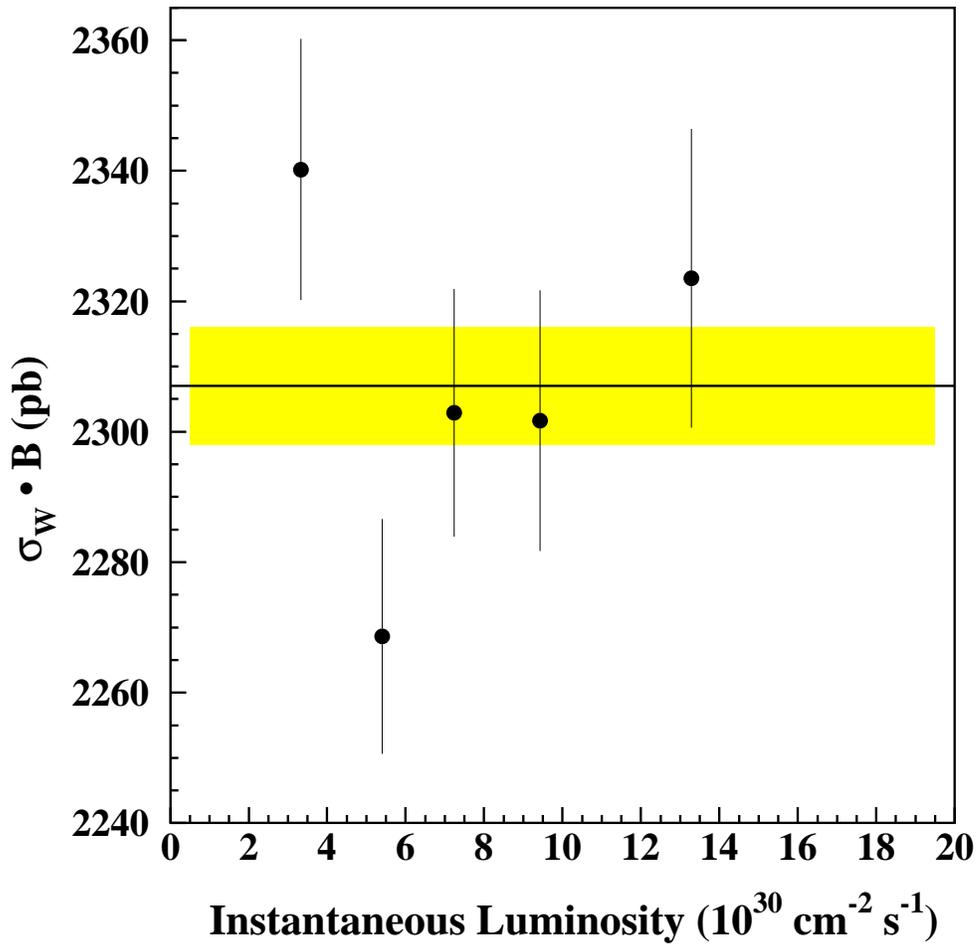,height=\figsize}}
        \caption{
The \wev\ cross section versus instantaneous luminosity.
The error bars are statistical only.
The solid line is the result from summing over all
instantaneous luminosities and the shaded band is the corresponding 
statistical uncertainty.  
}  \label{fig:wlum}
\end{figure}

\begin{figure}[hp!]
\center
\centerline{\psfig{figure=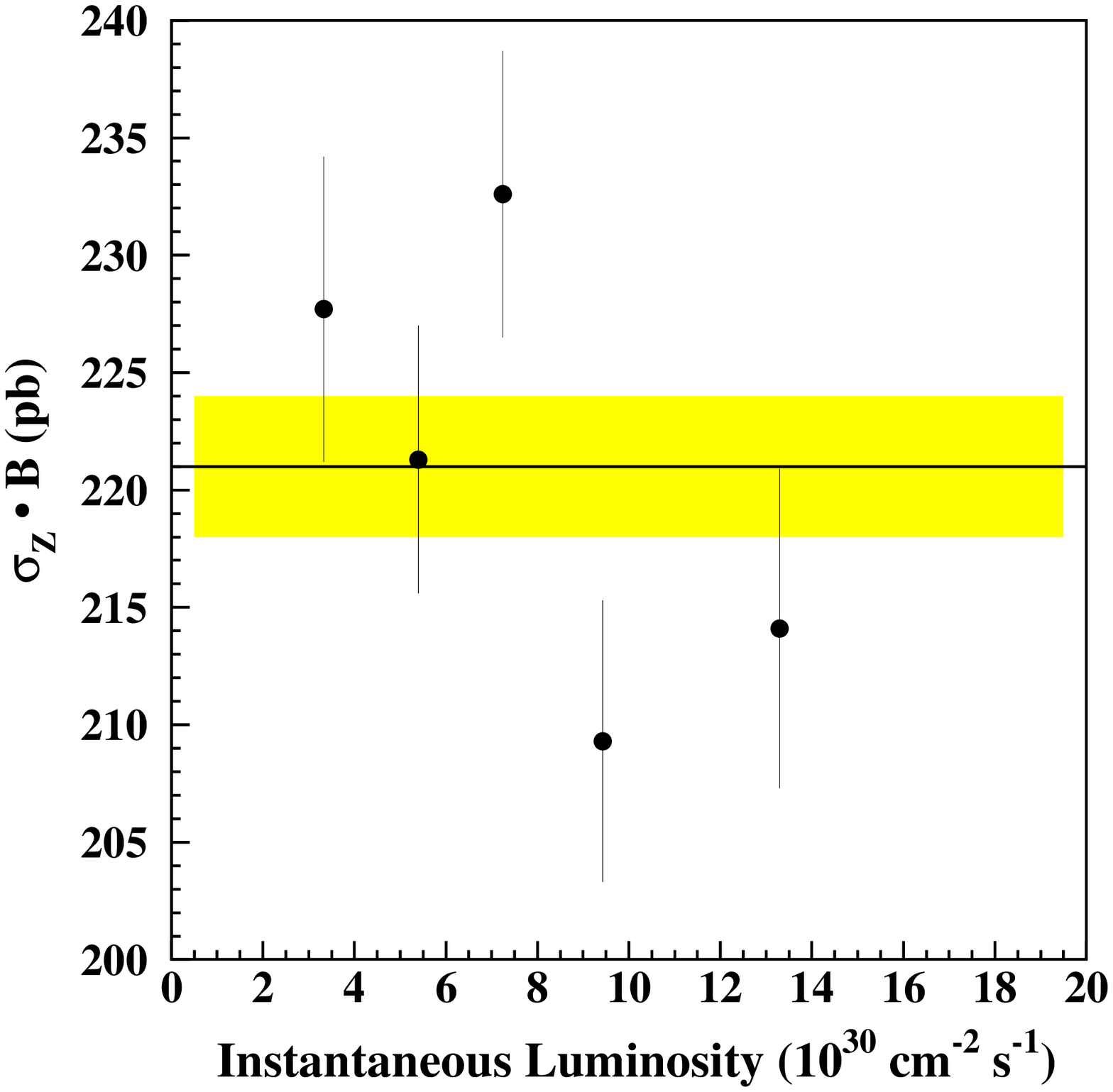,height=\figsize}}
        \caption{
The \zee\ cross section versus instantaneous luminosity.
The error bars are statistical only.
The solid line is the result from summing over all
instantaneous luminosities and the shaded band is the corresponding 
statistical uncertainty.
}  \label{fig:zlum}
\end{figure}

\begin{figure}[hp!]
\center
\centerline{\psfig{figure=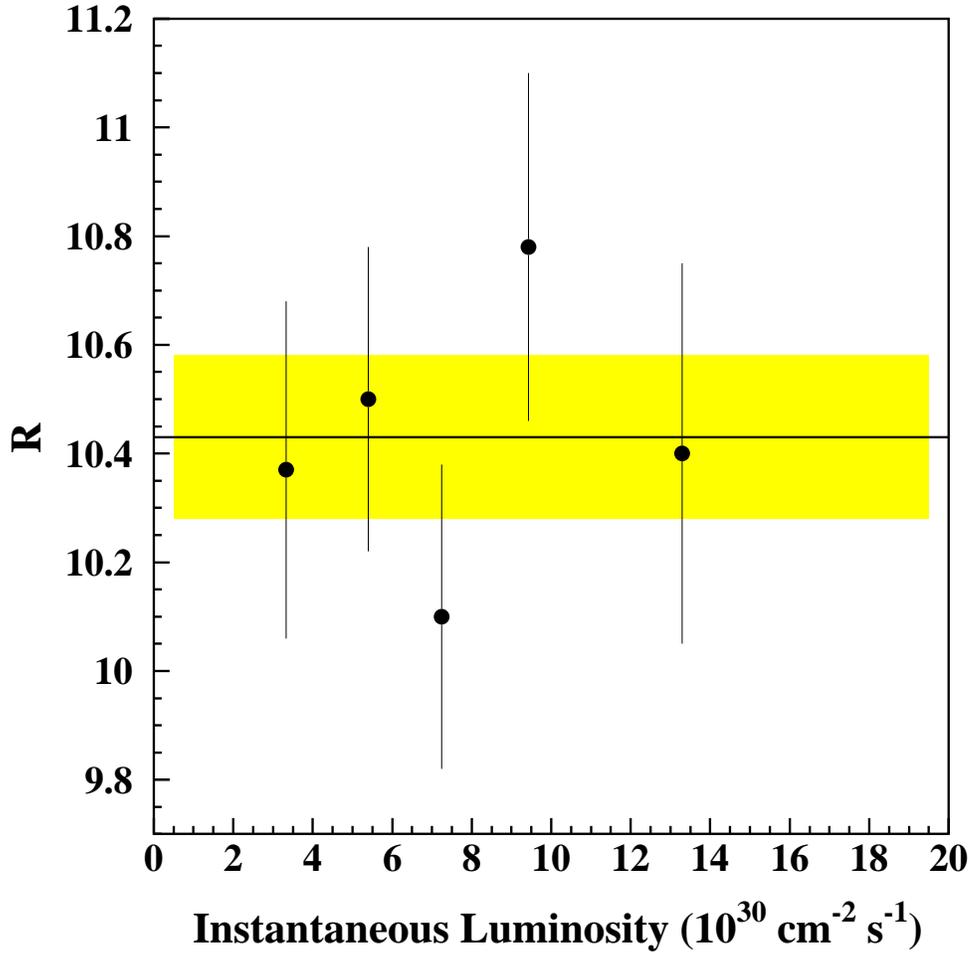,height=\figsize}}
        \caption{
The ratio \Reqn\ versus instantaneous luminosity.
The error bars are statistical only.
The solid line is the result from summing over all
instantaneous luminosities and the shaded band is the corresponding 
statistical uncertainty. 
}  \label{fig:rlum}
\end{figure}

To search for any dependences on luminosity, the data
are divided into five subsamples according to 
the value of the instantaneous luminosity when each event occurred
so that each subsample contains approximately one fifth of
the events.
The mean values of the instantaneous luminosity for each sample are 
3.33, 5.40, 7.24, 9.43, and 13.29 $\times 10^{30}$ cm$^{-2}$s$^{-1}$. 
For each subsample, the electron identification
efficiencies; the integrated luminosity; and the backgrounds
from multijet, $b$ quarks, and direct photons were re-calculated.
The electron identification efficiency for \wb\ boson events for the
highest luminosity bin is 17\% lower than that for the
lowest luminosity bin, and the multijet background is
2\% larger.
Figures~\ref{fig:wlum} and~\ref{fig:zlum} show the $W$ and $Z$ boson cross 
section, respectively, as a function of luminosity.
Figure~\ref{fig:rlum} shows  the ratio
of cross sections in the five bins of instantaneous luminosity.
The observed cross sections and their ratio do not appear
to depend on instantaneous luminosity, as the data are statistically
consistent with no luminosity dependence.

\section{\mbox{\boldmath \sigmaw\ at $\sqrts = 630$ \GeV }}
\label{sec:630}
We measure the \wev\ cross section using data from a short Tevatron run at a 
center-of-mass energy $\sqrt{s} = 630$ \GeV~\cite{kranethesis}. 
The integrated luminosity is calculated in the same way as for the 
$\sqrt{s} = 1800$ \GeV\ sample, except that we use values
of $\sigma_{sd}$, $\sigma_{dd}$, and $\sigma_{nd}$ which correspond to a
center-of-mass energy of 630 \GeV. These values are obtained by interpolating
between the measured values at $\sqrt{s} = 1800$ \GeV\ and at 
$\sqrt{s} = 546$ \GeV, and their uncertainties are dominated by the 
uncertainties at 546 \GeV. The luminosity calculation at $\sqrt{s} = 630$ 
\GeV\ is described in Ref.~\cite{lum630}. The integrated luminosity is  
\sslum\ $\pm$ \sslumerr\ \inb.
The cross section for inclusive \wb\ boson production at this center-of-mass
energy has previously been measured by the 
UA1~\cite{six30ua1} and UA2~\cite{six30ua2} collaborations.
We use the same \wev\ selection criteria
as was used  for the measurement at $\sqrt{s} = 1800$ \GeV, and
find a total of \ssnw\ \wev\ candidate events, \ssnwcc\ of which
have their electron in the \CC\ calorimeter. 

\begin{figure}[hp!]
\center
\centerline{\psfig{figure=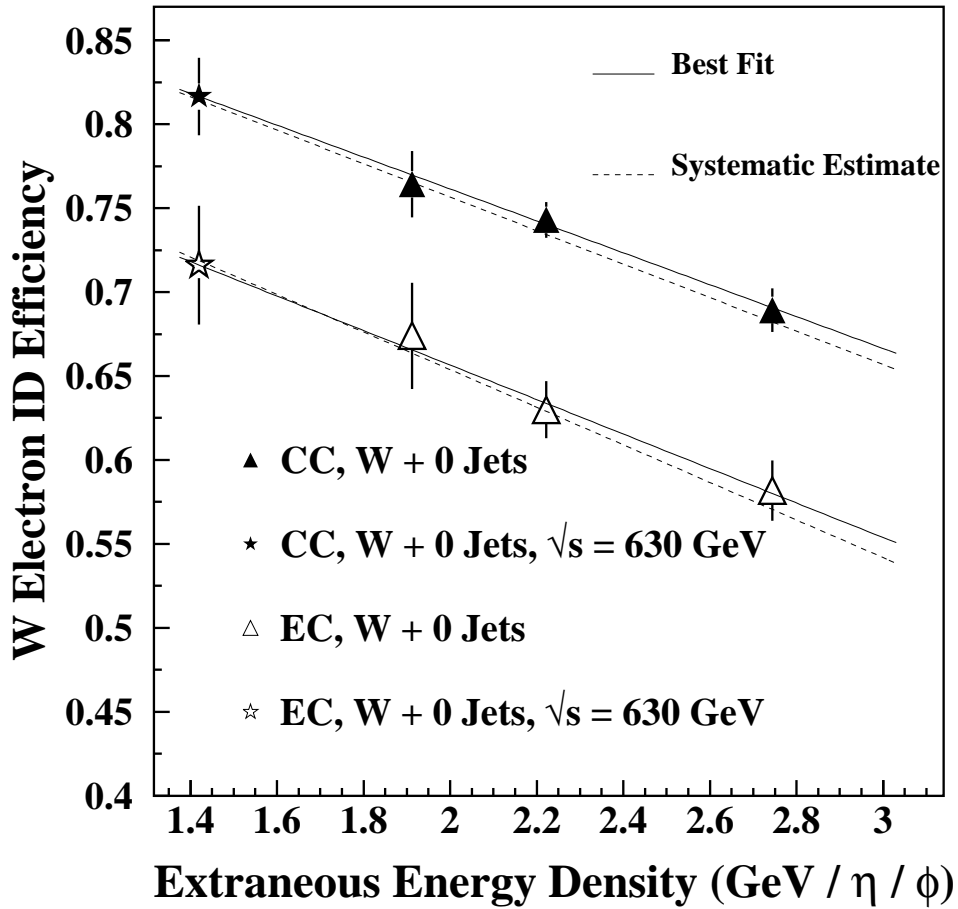,height=\figsize}}
\caption{
The efficiency for the electron identification criteria in the \CC\ and \EC. 
Each is measured as a function of the energy density in the event for events
containing no jets with \et $>$ 25 \GeV\ as evaluated using
the \zee\ sample taken at  $\sqrt{s}$=1800 \GeV .
The mean energy density of the sample taken at 
 $\sqrt{s} = 630$ \GeV\ is 1.3 \GeV$/ \eta / \phi$.
}
\label{fig:krane2}
\end{figure}

Since the 630 \GeV\ data sample contains very few \zee\ candidates
(approximately 10), we do not use this sample to obtain the 
electron identification efficiency.  Instead, we 
extrapolate the efficiency
from the 1800 \GeV\ sample.  The efficiency depends on the number of
jets in the \wev\ event and on the ambient energy in the event.
The \ssnw\ events from the \wev\ sample taken at $\sqrt{s} = 630$ \GeV\ 
contain no jets with $E_T > 25$ \GeV. 
Figure~\ref{fig:krane2} shows, for the \wev\ sample taken at 
$\sqrt{s} = 1800$ \GeV, the electron identification efficiency
for events without jets with $E_T > 25$ \GeV\ as a function
of the mean energy per unit of rapidity and per unit of azimuthal angle $\phi$.
The data sample taken at $\sqrt{s} = 630$ \GeV\ has a mean
energy density of 1.3 \GeV$/ \eta / \phi$, where $\phi$ is in radians.  
We fit the curve from the 1800 \GeV\ data to a first-order polynomial 
and use the fit to extrapolate to this energy density to obtain the 
efficiency of the electron identification requirements.  We obtain an electron
identification efficiency of 0.808 $\pm$ 0.024, where the uncertainty comes 
from the uncertainty in the fit.
The efficiency of the \lzero\ trigger for \wb\ boson events
relative to that for minimum bias events is also scaled from the 
result at 1800 \GeV. The \lzero\ trigger efficiency for minimum bias events at
1800 \GeV\ is \lzeight\ and at 630 \GeV\ is \lzsix.  The \wb\ boson efficiency
is scaled by the ratio of these two numbers and is
\sslzeff\ $\pm$ \sslzefferr.

The kinematic and fiducial acceptance is evaluated using the
same simulation as was used for the measurement at 1800 \GeV.
The fraction of
\wev\ events passing our fiducial and kinematic requirements at 
630 \GeV\ is 0.521 $\pm$ 0.013.

The background from multijets, $b$ quarks, and 
direct photons is calculated by scaling the 1800 \GeV\ result, using
$\displaystyle{ f_{QCD}^{630} = {P^{630}\over P^{1800}} \cdot
 {\sigma_j^{630} \over \sigma_w^{630}} \cdot
 {\sigma_w^{1800} \over \sigma_j^{1800}} } $,
where $P$ is the probability for a jet to fake an electron,
$\sigma_j$ is the cross section for jets with smeared
$p_T > 25$ \GeV, and $\sigma_w$ is the \wb\ boson cross section.
The \JETR~\cite{jetrad} program together with a parameterization
of the response of the detector has been shown to be
in good agreement with the data~\cite{someqcdpaper}.  Using
this program, we find
$\displaystyle{ {\sigma_j^{630} \over \sigma_w^{630}} \cdot 
{\sigma_w^{1800} \over \sigma_j^{1800}} = \ssrs\ \pm \ssrserr } $ and 
$\displaystyle{ {P^{630}\over P^{1800}} = \sspj }$, 
giving 
$f_{QCD}^{630}$= \fqcdsix\ $\pm$ \fqcdsixerr. The uncertainty is dominated by
the uncertainty in the jet cross section from \JETR.
The fraction of the candidates from the other sources of background
(\wtv , \zee , \ztt )
is assumed to scale in the same way as the signal with
center-of-mass energy.  Since \ssnzeight\ $Z$ boson events are expected to fake
$W$ boson events at 1800 \GeV, we expect \ssnzeight\ $\cdot$
\ssnw\ $/$ \wnum\ $=$ \ssnzsix\ $\pm$ \ssnzsixerr\ 
$Z$ boson events to pass the \wb\ boson selection at 630 \GeV.

\begin{figure}[hp!]
\center
\centerline{\psfig{figure=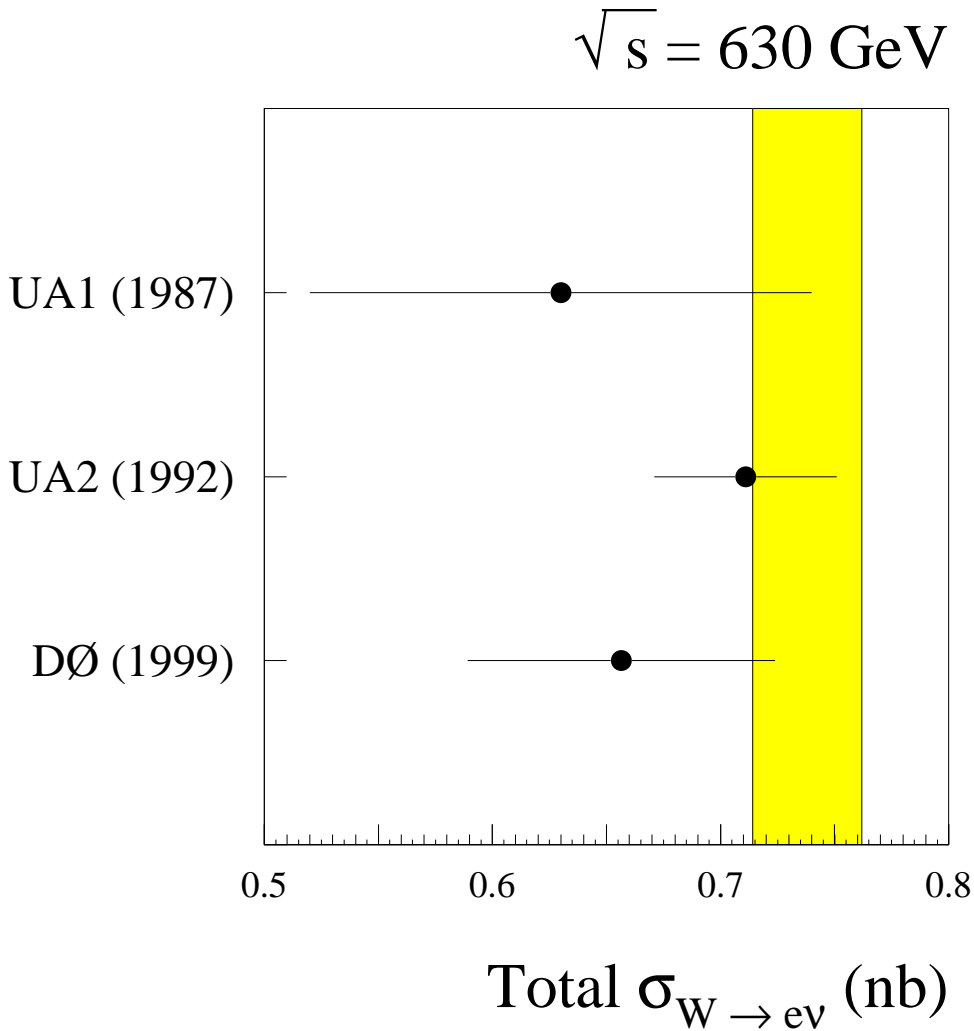,height=\figsize}}
\caption{
Measurements of the \wb\ boson inclusive cross section times electronic
branching fraction at a center-of-mass energy 
of 630 \GeV. Our result is in good agreement with previous measurements from 
the UA1 \protect\cite{six30ua1} and UA2 \protect\cite{six30ua2} collaborations.
The shaded band is a NLO prediction from the code of Ref.
\protect\cite{theoryrs}  with the CTEQ2M parton distribution functions, a $Z$ 
boson mass of 91.190 \GeV, a \wb\ boson mass of 80.23 \GeV, and 
$\sin{\theta_W}^2=$ 0.2259.
}
\label{fig:kraneua}
\end{figure}

Table~\ref{tab:w_table_630} and
Figs.~\ref{fig:kraneua} and~\ref{fig:crosssectionpic} summarize our result.
The result for \sigmaw\   is \wxssix\ $\pm$
\wxssixstat\ (stat) $\pm$ \wxssixsys\ (syst) \pb, 
where the systematic uncertainty includes a
3.0\% uncertainty in the integrated luminosity.

\begin{table}[h!]
\begin{center}
\caption{Values used in the \wev\ cross section measurement  at 630 \GeV. }
\medskip
\begin{tabular}{cccc}
&\sigmaw\      & \wxssix\ $\pm$ \wxssixerr\ \pb & \\
&              &                     & \\
\hline
&${N_{obs}^W}$ & \ssnw\ & \\
&${\epsilon_W}$ & \effsix\ $\pm$ \effsixerr & \\
&\lzero\ & \sslzeff\ $\pm$ \sslzefferr\ & \\
&${A_W}$ & \accsix\ $\pm$ \accsixerr & \\
&$Z$ boson background & \ssnzsix\ $\pm$ \ssnzsixerr\ & \\
&$(A_{Zee}^{W}+A_{Z\tau}^{W})/A_Z$ & \wacczinw\ $\pm$ \wacczinwerr & \\
&$f_{QCD}^W$ & \fqcdsix\ $\pm$ \fqcdsixerr & \\
&$A_{W\tau}^W/A_W$ & \wacctau\ $\pm$ \wacctauerr & \\
&$\cal L$ &  \sslum\ $\pm$ \sslumerr\ \inb\ & \\
\hline 
\end{tabular}
\label{tab:w_table_630}
\end{center}
\end{table}

%

\section{The Electronic Branching Fraction, Width, and
Invisible Width of the \mbox{\boldmath $W$} Boson }
\label{sec:results}
Using the results 
\sigmaw\ = \wxsec\ $\pm$ \wxstat\ $\pm$ \wxsyst\ $\pm$ \wxlum\ \pb ,
\sigmaz\ = \zxsec\ $\pm$ \zxstat\ $\pm$ \zxsyst\ $\pm$ \zxlum\ \pb , and 
${\cal R} = \rxsec\ \pm \rxstat\ \pm \rxsyst \pm \rxnlo$, 
we can determine the
electronic branching fraction of the \wb\ boson via
\begin{equation}
B( W \rightarrow e \nu ) = {\cal R} \cdot B(Z \rightarrow ee) \cdot
\frac{\sigma_Z}{\sigma_W}
\end{equation}
Using $B( Z \rightarrow ee)$ = 0.03367 $\pm$ 0.00006~\cite{lepzee} and
$\sigma_W / \sigma_Z$ = 3.29 $\pm$ 0.03~\cite{theoryrs}, we get
$B(W\rightarrow e \nu)$ = \brwev\ $\pm$ \brstat\ (stat)
$\pm$ \brsyst\ (syst) $\pm$ \brthy\ (other) $\pm$ \brnlo\ (NLO), where the 
next-to-last source of 
uncertainty comes from uncertainties in $B( Z \rightarrow ee)$ and 
in $\sigma_W / \sigma_Z$.
The standard model prediction is $B(W\rightarrow e \nu) =
0.1084 \pm 0.0002$.
Assuming  the standard model prediction for the electronic
partial width (0.2270 $\pm$ 0.0011 \GeV~\cite{theorywidth}), 
we can calculate the $W$ boson width
$\Gamma_W = \Gamma_W^e / B(W \rightarrow e \nu)$ as
\gw\ $\pm$ \gwstat\ (stat) $\pm$ \gwsyst\ (syst) $\pm$ \gwthy\ (other) 
$\pm$ \gwnlo\ (NLO) \GeV, 
to be compared with the standard model prediction of
$\Gamma_W$ = 2.094 $\pm$ 0.006 \GeV~\cite{theorywidth}.
The difference between our measured value and the standard model prediction, 
which is the width for the \wb\ boson to decay to final states
other than the two lightest quark doublets and the three
lepton doublets, is thus 0.036 $\pm$ 0.060 \GeV. This is
consistent with zero within uncertainties, so we set a 95\% confidence level 
upper limit on the \wb\ boson width to non-standard-model final states 
(``invisible width").
Assuming the uncertainty is Gaussian, removing the unphysical region
where the invisible width is negative, and integrating to 95\% of the remaining
area, we set a 95\% confidence level upper limit
on the invisible partial width of the \wb\ boson of \gwinvgev\ \GeV.

We combine our Run 1b (1994--1995) result with the \Dzero\ results from 
Run 1a (1992--1993)~\cite{d01aR} for $\cal R$. 
Table~\ref{tab:compare1a} compares the two measurements.
\begin{table}[ht!]
\begin{center}
\caption{Comparison of the current Run 1b (1994--1995) measurement to the 
Run 1a (1992--1993) measurement. }
\medskip
\begin{tabular}{|l|c|c|c|}\hline\hline
Data Period & $\cal R$ & Correlated uncertainty  & Uncorrelated uncertainty \\
& & & \\
\hline
1a,electron (\luma\ \ipb) & \ra & \raerrcorr & \raerruncorr \\
1a,muon (\lumamu\ \ipb) & \ramu & 0 & \ramuerruncorr \\
1b,electron (\lumb\ \ipb) & \rxsec & \rberrcorr & \rberruncorr \\
\hline\hline
\end{tabular}
\label{tab:compare1a}
\end{center}
\end{table}
Because most of the systematic uncertainties in the Run 1a
measurement in the electron channel were dominated by the statistics of the 
sample used to evaluate the uncertainty, the 1a and 1b measurements in the
electron channel
are mostly uncorrelated.  Only the acceptance, the
Drell-Yan correction, and the NLO uncertainties are correlated (we have
added the same 1\% NLO uncertainty to the 1a result). 
The measurements in the muon and electron channels are uncorrelated. With 
this assumption,
we get ${\cal R} = \rcomb \pm \rcomberr$, $\Gamma_W = \gwcomb \pm 
\gwcomberr\ $ \GeV , 
and a 95\%
confidence level upper limit on the invisible width of \gwinvcomb\ \GeV .
Table~\ref{tab:results} summarizes our results.
\begin{table}[h!]
\begin{center}
\caption{Results.}
\medskip
\begin{tabular}{|l|c|c|}\hline
 & 1b   & 1a+1b combined  \\
 & (\lumb\ \ipb) & (\luma\ + \lumamu\ + \lumb\ \ipb) \\
\hline
Ratio $\cal R$  & \rxsec\ $\pm$ \rerr  & \rcomb\ $\pm$ \rcomberr \\
$B(W \rightarrow e\nu)$ & \brwev\ $\pm$ \brerr & \brcomb\ $\pm$ \brcomberr \\
$\Gamma_{W}$ & \gw\ $\pm$ \gwerr\ \GeV  & \gwcomb\ $\pm$ \gwcomberr\ \GeV \\
95\% C.L. upper limit $\Gamma_{W}^{inv}$ & \gwinvgev\ \GeV & \gwinvcomb\ 
\GeV \\
\hline
\end{tabular}
\label{tab:results}
\end{center}
\end{table}



\section{Conclusions}
\label{sec:conclusions}
We have presented new measurements of \sigmaw\ and \sigmaz\ using
\lumb\ \ipb\ of data. We determine
\sigmaw\ $= \wxsec \pm \werr$ \pb\ and \sigmaz\ $= \zxsec \pm \zerr$ \pb. 
The uncertainty in these measurements is dominated by the luminosity 
uncertainty. From these measurements, we have derived the ratio 
${\cal R} = \rxsec \pm \rerr $  
and a new indirect measurement of the total \wb\ boson width, 
$\Gamma_{W} = \gw \pm \gwerr$ \GeV. We obtain a 95\%
confidence level upper limit on the invisible \wb\ boson width 
of \gwinvmev\ \MeV. 
Combining these results with those from the 1992--1993 
run~\cite{d01aR}, we determine ${\cal R} = \rcomb \pm \rcomberr$, 
$\Gamma_W = \gwcomb \pm \gwcomberr\ $ \GeV , 
and a 95\%
confidence level upper limit on the invisible width of \gwinvcomb\ \GeV .

\section*{Acknowledgements}

%
We thank Ulrich Baur for helpful discussions and calculations concerning 
electroweak radiative corrections and diffractive \wb\ and \zb\ boson 
production.
We also thank John Collins, James Stirling, and Willy van Neerven for 
discussions about diffractive $W$ and $Z$ boson production.
We thank the Fermilab and collaborating institution staffs for
contributions to this work and acknowledge support from the 
Department of Energy and National Science Foundation (USA),  
Commissariat  \` a L'Energie Atomique (France), 
Ministry for Science and Technology and Ministry for Atomic 
   Energy (Russia),
CAPES and CNPq (Brazil),
Departments of Atomic Energy and Science and Education (India),
Colciencias (Colombia),
CONACyT (Mexico),
Ministry of Education and KOSEF (Korea),
and CONICET and UBACyT (Argentina).
%



\clearpage


\begin{thebibliography}{99}
\bibitem{corr_example}
An example may help make this cancelation more intuitive.   
Suppose that most of the time, the probability
that one of the electrons in a $Z$ boson event passes the tracking cuts
is uncorrelated with the probability that the other electron passes.
If, for certain runs, a detector malfunction were to cause  
all tracking to fail, the probability that one electron fails (100\%) is 100\%
correlated with the probability that the other fails (100\%). 
Since all $Z$ and $W$ boson events that occur during this hardware failure 
would be lost, the loss cancels in the ratio.
\bibitem{lum_comment}
Luminosity measurements are stored periodically (every minute) and then 
integrated over the live running time. This one-minute period is much shorter 
than the time it takes for the luminosity to change significantly.
\end{thebibliography}
\end{document}
%